\documentclass[12pt]{amsart}
\usepackage[margin=1in,letterpaper,portrait]{geometry}
\usepackage{amsthm,amssymb,
enumitem,mathtools,nicefrac}
\usepackage{hyperref}

\usepackage{quoting}
\quotingsetup{vskip=0pt}

\usepackage{enumitem}

\usepackage{tikz}
\usetikzlibrary{shapes.geometric,calc}

\usepackage{cite}

\theoremstyle{plain}
\newtheorem{thm}{Theorem}

\newtheorem{lemma}[thm]{Lemma}
\newtheorem{defn}[thm]{Definition}

\theoremstyle{remark}
\newtheorem*{flaw-ex}{Example}

\renewcommand{\sp}{{\sf sp}}
\newcommand{\cut}{{\sf cut}}

\newcommand{\coastline}{\medskip \noindent {\bf Issue A: Physical geography.}
{\em Districts with edges defined by natural features, such as coastlines, are heavily penalized by contour-based scores, whereas 
physical features are often good choices for district boundaries.  \vspace{.08in}}} 

\newcommand{\stability}{\medskip \noindent {\bf Issue B: Resolution instability.} 
{\em Varying map resolution can have a dramatic impact on  contour-based  scores, even though resolution is functionally independent of the district definition.  \vspace{.08in}}}

\newcommand{\coord}{\medskip \noindent {\bf Issue C: Coordinate dependence.}
{\em The choice of map projection and coordinate system, while fully independent of district definition, can impose drastic changes on the contour-based scores.  \vspace{.08in}}}

\newcommand{\emptyspace}{\medskip \noindent {\bf Issue D: Areal emphasis.} 
{\em Contour-based scores emphasize land area, despite the fact that the electoral impact of districts is purely population-based. \vspace{.08in}}}

\newcommand{\PP}{{\sf PP}}
\newcommand{\area}{{\sf area}}
\newcommand{\perim}{{\sf perim}}

\newcommand{\sdot}{\!\cdot\!}

\begin{document}

\title{Discrete geometry for electoral geography}

\author[Duchin]{Moon Duchin}
\address{Department of Mathematics, Tufts University, Medford, MA 02155}
\email{moon.duchin@tufts.edu}

\author[Tenner]{Bridget Eileen Tenner}
\address{Department of Mathematical Sciences, DePaul University, Chicago, IL 60614}
\email{bridget@math.depaul.edu}

\thanks{MD was supported by NSF DMS-1255442 and DMS-2005512. BT was partially supported by Simons Foundation Collaboration Grant for Mathematicians 277603, a DePaul University Faculty Summer Research Grant, and NSF DMS-2054436.}


\maketitle

\begin{abstract}
``Compactness,'' or the use of shape as a proxy for fairness, has been a long-running theme in the scrutiny of electoral districts; badly-shaped districts are often flagged as examples of the abuse of power known as {\em gerrymandering}.
The most popular compactness metrics in the redistricting literature belong to a class of scores that we call {\em contour-based}, making heavy use of area and perimeter. 
This entire class of district scores has some common drawbacks, outlined here.
We make the case for {\em discrete} shape scores and offer two promising ideas: a {\em cut score} and a {\em spanning tree score}.  We use recent United States redistricting history as a source of examples.

No shape metric can work alone as a seal of fairness, but we argue that discrete metrics are better aligned both with the grounding of the redistricting problem in geography and with the computational tools that have recently gained 
significant traction in the courtroom.  \\

\end{abstract}


\section{Introduction}

A variety of elections in the United States---for the House of Representatives, state legislatures, city councils, school boards, and more---are conducted by partitioning a local area into geographically-delimited \emph{districts} and selecting one winner per district via a plurality 
election. A suitable partition of the locality is called a \emph{districting plan}, or sometimes just a \emph{plan}, and the act of revising it is called \emph{redistricting}. 
States may have their own guidelines governing the redistricting process, often including specific requirements for valid plans, and the procedures and outcomes are the subject of considerable debate and legal scrutiny.
Both mathematicians and geographers are professionally interested in boundaries, making for natural allies to study the intersection of shape, territory, and representation.  
We set out here to  re-think the ``boundary work'' of electoral terrain \cite{gieryn}.

There are two main principles commonly applied to the shape (that is, the geometric form) of districts:  jurisdictions should be cut into pieces that are ``contiguous'' and ``compact.'' 
The first of these criteria, \emph{contiguity}, refers to topological connectedness: a district should be a single connected component, not multiple separated components. This is a widespread and mostly
uncontroversial requirement for districting plans.\footnote{Note that while contiguity seems largely unambiguous, it is sometimes achieved by connective tissue along a highway or through water. For example, after the 2000 Census, Illinois's 4th Congressional District had been constructed 
using a stretch of Interstate 294 to connect its northern and southern components, in a shape often described as ``earmuffs.'' 
Many states need to interpret contiguity across water, which is not always handled in a clear or consistent fashion (and see \cite{Alaska} for an example of surprising impacts).} 
In contrast with that clarity, {\em compactness} gestures at the idea that shape should be somehow reasonable rather than eccentric, but this is rarely defined precisely, if at all.  Even on such unsteady footing, 
the notion is critical to any discussion of redistricting (and, in particular, to any discussion of abusive districting practices broadly
known as {\em gerrymandering}) because compactness appears as an 
explicit requirement in many states and is nationally recognized as a traditional districting principle.\footnote{See \href{https://redistricting.lls.edu/where-state.php}{\tt redistricting.lls.edu/where-state.php} (accessed July 26, 2023), which describes specific compactness rules of some kind in more than 30 of the 50 states.} 
And in the biggest redistricting court case of this census cycle so far---a complaint on behalf of Black voters about the congressional district boundaries in Alabama---the Supreme Court's 
decision to invalidate the state's map made heavy reference to the plaintiffs' demonstration plans being ``reasonably configured,'' in large part because their compactness scores were 
``generally better on average than'' the state's plan, and they contained no ``bizarre shapes, or any other obvious irregularities.''\footnote{{\em Merrill v Milligan} (2023).
Disclosure: MD served as an expert in litigation in several  court cases mentioned in this article, particularly {\em LWV v. Pennsylvania} (2018) and {\em Allen v. Milligan} (2023).
}

To date, more than thirty possible definitions of compactness as a shape quality metric have been proposed in the political science literature (see \cite{altman, gerrybook, NGCH} and references).
The purpose of this paper is to call attention to a shared feature in nearly all of the popular definitions of 
compactness scores:  they dissolve the geographical units of districts and represent each district as a single region enclosed by a contour on a projected map.  Numerical scores are then based on measurements made relative to the contours.
This makes all of the standard definitions of compactness
susceptible to a common set of drawbacks, which undermine the extent to which the definitions can be made precise and meaningful.  
More broadly, passing up units for contours risks making a category error.  Underlying other data formats, districting plans are defined and communicated as census block assignments, so there is a basic sense in which the problem is fundamentally discrete.

Geographers---especially but not only those who work with GIS---are intimately familiar with this tension.  
Working with spatial data involves both geographical units and coordinate mapping; geography more generally contends
with the interplay between the discrete and the continuous, and with problems of scale and zoning that come from that interplay, all the time and fundamentally.  Managing the confounding effects of the choice of units is a central theme in the field, going by the name of the Modifiable Areal Unit Problem, or MAUP, most associated with the work of Openshaw \cite{Openshaw1,Openshaw2} but anticipated  in  \cite{MAUP2} and developed in \cite{MAUP1,MAUP3} and in more recent work like \cite{MAUP4,MAUP5}.
A discrete approach to shape can make  fundamental use of the geographical units and the demographic network structure that are at the heart of the redistricting problem, confronting the redistricting MAUP head-on. 
Furthermore, this discrete approach is well adapted to the mathematical and computational tools that are increasingly prominent in legal settings.

\subsection{Outline}

We give a brief background on compactness and census/electoral geography in \S\ref{sec:cpct-bg}, survey the standard contour-based compactness scores and their use in courts and by redistricting bodies in \S\ref{sec:contour}, then present and discuss  problems that contour scores face in \S\ref{sec:issues}.  Graph formalism is introduced in \S\ref{sec:discrete} and two scores---a cut score measuring partition efficiency and a spanning tree score measuring clustering---are defined. Finally,
\S\ref{sec:assess} provides tools for interpretation and for assessment, both abstractly and empirically.  

The use of graph partitions to model redistricting is hardly new; in particular, it is at the center of every algorithmic approach to generating districts, now held up as a promising tool for understanding redistricting in a set of opinions that collectively include every sitting U.S. Supreme Court justice as a signatory.\footnote{The main opinion, the concurring opinion, and the dissenting opinions in {\em Milligan} all describe potentially valuable roles for algorithmic district sampling, with various degrees of enthusiasm.  See \cite[Ch 16]{gerrybook} for a historical survey of algorithmic approaches.}   Computational redistricting has been a long-standing dream since at least the 1960s and has exploded as a practical reality 
in U.S.-based research since the 2010 Decennial Census. The cut and spanning tree scores presented in \S\ref{sec:discrete} are extremely natural in the graph setting and tie into a large body of mathematical literature, including the study of so-called {\em community structure} in networks and efficient algorithms in scientific computing.  Our aim is to give these topics a unified and simple treatment  that dovetails with the normative and substantive grounding of compactness in representational and geographical terms.

\subsection*{Acknowledgements}
This conversation began during the Geometry of Redistricting Workshop held at Tufts University in August 2017. We are grateful to the speakers and participants.  Special thanks  to Assaf Bar-Natan, Richard Barnes, Daryl DeFord, Seth Drew, Adriana Rogers, Parker Rule, and Alejandro Velez-Arce for their excellent work with data and computation, and for illuminating conversations.  We relied heavily on their data and materials in our thinking and writing about this topic.  We also thank Heather Rosenfeld, Lee Hachadoorian, Ruth Buck, Benjamin Forest, Peter Winkler, and Justin Solomon for overlapping collaborations, feedback, and conversations.

\section{Compactness and electoral geography}\label{sec:cpct-bg}

\subsection{Introducing compactness}

The political relevance of requiring districts to be reasonably shaped, and not unnecessarily  
elongated or twisting, can be defended in several ways:  shape assessment detects signals of manipulation, imposes checks on power, and promotes cognizable and functionally interconnected districts.

First and foremost, geometric eccentricity can 
signal a districting plan that has been  engineered to produce an 
extreme outcome,  for instance by exploiting demographics and geography in order to maximize representation for one group at the expense of another.  
A mapmaker can tilt outcomes by {\em packing} the out-group into a small number of districts, with wastefully high vote share in those districts, and {\em cracking}
their leftover population by  dispersing it, thus diluting those voters' influence.  
Either strategy can induce (indeed, might necessitate) distended district shapes in order to unite non-proximal groups to create a carefully composed district.
In fact, it has been convincingly argued that the contorted appearance of a district can have an ``expressive harm,'' communicating to voters that fundamental criteria were subordinated achieve to other goals, such as manipulating the racial composition of the district \cite{bizarre}.  This idea, that appearances can be directly harmful, amounts to a defense of the ``eyeball test'' for district shape.

A second, related argument for shape guidelines is that any limitation placed on districters
is a healthy check on their power.  A third kind of argument, this time positively framed, argues that being more compact 
should mean that districts represent chunks of territory that have a social or infrastructural cohesion, and can be traveled efficiently.
This ties into the idea of {\em cognizable} districts: 
the territory of a district should be distinguishable by its residents and its representative, and should correspond reasonably with the structure of towns and counties, making it easy to describe.\footnote{This cognizability principle was examined in some depth by political scientist Bernie Grofman:  ``By `cognizability,' I mean the ability to characterize the district boundaries in a manner that can be readily communicated to ordinary citizens of the district in commonsense terms based on geographical referents'' \cite{lombardi}.  
The original Gerry-mander of 1812 was singled out  for flouting this kind of easy delineation through suspiciously complicated selection of towns and its unnecessary division of counties \cite{OG}.
See also Benjamin Forest, who looks at the U.S. Supreme Court rhetoric from  their ``Shaw line'' of rulings, which focused on non-compactness and racial gerrymandering \cite{forest}.  In Forest's view, the justices held that compact plans must respect the political regionalization of a state (into counties and municipalities) as well as its physical geography.
}  
Clear and easily communicated boundaries might enable the representative to better understand their electorate; conversely, constituents need to be able to identify and contact their representative \cite{zipcodes}.

Ease of transit and communication is sometimes bundled into a notion of ``functional compactness'' together with other traits such as community composition.  The notion of {\em communities of interest}---COIs for short---is about factors such as social organization, identity, and economics  combining to create geographical areas that may not coincide with administrative boundaries, but which are still meaningful and relevant to political representation.  
Districts are regarded as more successful when COIs are held intact, and can be faulted for splitting them, or for uniting them improperly.  
Though this at first sounds completely independent of district shape per se, legal decisions have sometimes blurred the lines.\footnote{To see the intertwining of shape and community structure, consider this passage from Justice Kennedy's majority opinion in {\em LULAC v Perry} (2006):
``The enormous geographical distances separating the two communities, coupled with the disparate needs and interests of these populations---not either factor alone---renders District 25 noncompact for \S2 purposes.''  So the sins of this Texas congressional district are at the same time based on geometry and on a lack of social/demographic harmony.  More examples of this entwinement of shape and community talk are traced through earlier court decisions in \cite{forest}.}

So in sum, a good compactness definition should flag geographically complicated boundaries as bad; should meaningfully constrain the space of allowable plans; should tend to label easily described districts as good; and should comport with a general visual sense of a shape's simplicity on a map.  Most ambitiously, it might also tie, in some way, to notions of community.
An elementary discussion of compactness metrics can be found in \cite[Ch 1]{gerrybook}; a review of regionalization perspectives on redistricting appears in \cite[Ch 11]{gerrybook}; and a brief overview of communities of interest law and practice is offered in \cite[Ch 12]{gerrybook}. 

Experts routinely testify that there are multiple compactness scores with no single best choice of score and no bright-line threshold of permissibility; furthermore, they are hard to compare and contextualize.  So what accounts for their enduring popularity?  Scores are appealing to courts because they put some quantitative meat on the bones of intuitive visual assessment.  This means that they offer some (apparent) concreteness in a domain that is notorious for the lack of agreed-on standards.  Furthermore, as is very often the case, technology is a key part of the story:  we cannot understand the uptake of particular compactness scores without investigating their ease of use in the dominant software packages.  Expert work involving measures of compactness draws on either commercial software such as Maptitude for Redistricting, on free software such as Dave's Redistricting App (\href{https://davesredistricting.org}{\tt davesredistricting.org}), or on custom functions built in GIS or in geospatial packages in Python or R.  
Maptitude, in particular, is the dominant enterprise package that legislatures and their consultants use to draw the lines; states, counties, and cities frequently release their districting plans together with a suite of Maptitude reports.    Because the software plays such a central role, we will cite Maptitude functionality as a recurring theme below.

\subsection{Districts and their building blocks}

Though political jurisdictions, even states themselves, have textual legal definitions, the usable formats for 
communicating those definitions are usually based on technological representations found in a GIS shapefile.
Shapefiles store a definition of each unit as a polygon, possibly with many thousands
of vertices.  In U.S. redistricting, the canonical shapefiles are data products released by the Census Bureau.
The Census Bureau releases an updated  ``vintage'' of its most precise shapefiles---called \emph{TIGER/Line Shapefiles} 
\cite{tigerline}---every year, with a special release for congressional districts once new boundaries have been enacted in law
after each decennial data drop.  It also releases Cartographic Boundary Shapefiles of districts  \cite{cartographic}, 
which are intended for the purpose of map-drawing rather than definition, and are generally clipped to land (see Figure~\ref{fig:AL-1}). These Cartographic maps are prepared for every Congress, 
at three levels of resolution, discussed further below.

U.S. Census geography is organized in a hierarchy that begins with finest units called {\em blocks}, 
nesting into larger units called {\em block groups}, which in  turn nest into {\em tracts}, then counties and states.\footnote{There were over 11 million blocks in the 2010 Census, with an average population of about 28.}
For each level in the main census hierarchy (also called the ``central spine''), the geographical units in that category partition the state that they belong to, meaning that 
 the entire territory of the state (land and water) is covered by the census units at that given scale, and furthermore that those units are disjoint from each other, except along their borders.  In mathematical language, the units {\em tile} the state at each level, and smaller units nest into larger ones.

Not every relevant geography is defined by the Bureau.
{\em Precincts} are the state or local administrative units of geography in which elections are conducted and vote results are reported.
Their boundaries are typically controlled by local officials, and they can change at erratic intervals.  
Because of their importance for elections, the Census Bureau attempts to capture them in an approximate snapshot every ten years known as  {\em voting tabulation districts} or VTDs. These are  made of whole census blocks, but do not respect block groups or any other units in the central spine.
Sometimes it is appropriate to model the redistricting problem with precinct assignments, as numerous states have a strong preference or even a requirement for keeping precincts intact in particular kinds of districts, largely because election administrators struggle to deliver the correct ballots when precincts are split.

But blocks, the finest level of census detail, will remain the principal atoms for redistricting.  Districts form an off-spine partition of the state, made after the decennial release using census blocks, while their lines frequently cut  across block groups, VTDs, and tracts.
Since blocks nest inside of all other census geographies,  other units' populations 
are calculated by aggregation from the blocks.  
And indeed the call to release data on blocks is explicitly for redistricting; Public Law 94-171 was passed in 1975 and mandated a decennial release known as the Redistricting Data, sometimes itself referred to as the PL94-171 data.  This tabular data is  composed of counts of residents by race, ethnicity, and voting age status for each block \cite{PL94}.
These blocks are then used to fine-tune populations on districts, with a common practice of balancing Congressional plans so that the top-to-bottom deviation is {\em one person}---which would be essentially impossible with larger units than blocks.\footnote{\emph{Reynolds v Sims} (1964) gave the general slogan ``One Person, One Vote,'' which by common
practice has come to require the near-equalization of census population across congressional districts, with a few more percentage points of slack at legislative levels and below. Congressional plans are often ``zero-balanced," while legislative districts are typically allowed top-to-bottom deviation up to ten percent of ideal size. For actual population balances in enacted districting plans, see \cite{NCSL}.}
This makes census blocks  the standard ``pixels'' of redistricting and explains why block assignment files are a common data format, more succinct but just as precise as a shapefile of districts.

\section{Polsby-Popper and other contour-based scores}\label{sec:contour}

\begin{quote} ``Compactness,'' unlike contiguity, is a continuous concept that concerns the geographical shapes of districts. There is no bright line test that determines whether a district is or is not compact, but districts may be considered more or less compact. While numerous quantitative measures of compactness have been proposed for this purpose, the two measures that are now referenced the most are a dispersion measure known as the Reock measure and a perimeter measure known as the Polsby-Po[p]per measure.\end{quote}
\begin{flushright}{--Expert report of Dick Engstrom, \emph{Martinez v Bush} (2002)}\end{flushright}

The most commonly cited compactness metric in litigation is the {\em Polsby-Popper score}. The motivating idea for Polsby-Popper and its cousins is that a ``compact'' region should have large area relative to its perimeter.
This is an \emph{isoperimetric} score, because it creates a ranking among regions with a given 
(``iso'' = same) perimeter.

\subsection{Perimeter versus Area}

It has been known (or guessed) since antiquity that
\begin{center}
\emph{circles have the most area among all shapes with a given perimeter.}
\end{center}
In other words, all shapes satisfy $0\le \frac{4\pi A}{P^2}\le 1$, where $A$ stands for area and $P$ stands for perimeter.  Examples of this phenomenon appear in Figure~\ref{fig:isoperimetry}. Depending on the scope of the statement 
(i.e., on the generality of what counts as a ``shape''), the first rigorous proof  can be credited to Jakob Steiner in the 1830s.  
Here is a modern statement of the phenomenon.

\begin{figure}[htbp]
\centering
\begin{tikzpicture}[scale=3] 

\begin{scope}[xshift=-.7cm,yshift=.125cm]
\draw (0,0) rectangle (.48,.02);
\end{scope}

\draw (0,0) -- (1/4,0) -- (1/4,1/16) -- (1/16,1/16) -- (1/16,1/4) -- (0,1/4) -- (0,0);

\begin{scope}[xshift=.5cm] 
\draw (0,0) -- (.333,0) -- (.167,.289) -- (0,0);
\end{scope}

\begin{scope}[xshift=1.1cm] 
\draw (0,0) rectangle (.25,.25);
\end{scope}

\begin{scope}[xshift=1.8cm] 
\draw (0,.125) circle (.159);
\end{scope}
\end{tikzpicture}
\caption{Five regions with the same perimeter are shown from left to right in order of increasing area. The region with largest possible area relative to a fixed perimeter is the circle, and is deemed the most ``compact''
by Polsby-Popper scoring.
Reading left to right, the Polsby-Popper scores are roughly .12, .34, .60, .79, and exactly 1.}\label{fig:isoperimetry}
\end{figure}

\begin{thm}[Isoperimetric Theorem]\label{thm:isoper}
Let $\Omega$ be a bounded open subset of the Euclidean plane $\mathbb R^2$ whose boundary $\partial\Omega$ is a rectifiable curve.
Then the Lebesgue measure $m$ and the length $\ell$ are related by the inequality
$4\pi \sdot m(\Omega)\le \ell(\partial \Omega)^2$,
with equality if and only if $\Omega$ is a disk and $\partial\Omega$ is a circle.
That is, all $\Omega$ satisfy $$4\pi \frac{\area(\Omega)}{\perim(\Omega)^2}\le 1.$$
\end{thm}

In this generality, Theorem~\ref{thm:isoper} has a short and elegant proof using the Brunn-Minkowski inequality \cite{shakarchi}. Despite the long history of this fact (see especially \cite{Blasjo} for an excellent guided tour),  a 1991 article by law scholars Polsby and Popper 
led to their names being attached to the associated metric in political science \cite{PP}.  

\begin{defn}\label{defn:pp}
The \emph{Polsby-Popper score} of a district $\Omega$ is
$$\PP(\Omega):=4\pi\sdot \frac{\area(\Omega)}{\perim(\Omega)^2}.$$
\end{defn}

Equivalently, $\PP(\Omega)$ can be defined as the ratio of the area of $\Omega$
to the area of the circle whose circumference is equal to the perimeter of $\Omega$.  (To see this, set $2\pi r = \perim$ to define the circle, solve for the radius $r$, and compute $\pi r^2$.) 

The Polsby-Popper score of a districting {\em plan} is not defined in the literature. 
Legal and administrative reports often include the mean, maximum, minimum, and other such statistics for the 
set of scores over the districts in a plan, likely because this is made easy in the standard commercial software  (see Figure~\ref{fig:maptitude}).

The Polsby-Popper score is highly sensitive to elongations of the boundary.
Shapes with skinny necks, long spurs, or whose boundary winds in a complex manner  will have much less area than 
could have been enclosed by the same boundary length around a ``plumper'' shape. 
Higher Polsby-Popper scores are therefore termed more compact, and are thought to be preferable to lower ones. 

By Theorem~\ref{thm:isoper}, the score satisfies $0\le \PP(\Omega)\le 1$ for all shapes, with $\PP(\Omega)=1$
realized only when $\Omega$ is a circle. The squaring of the perimeter in the denominator of the Polsby-Popper score also serves to make the units of measurement cancel out, so that the score is (theoretically) scale-invariant. In other words, if one were to dilate an entire region by a factor of $k$, its Polsby-Popper score would 
not register the change.  Thus this metric is said to measure something about the shape, and not the size, of a district.  Polsby-Popper scores have been cited in hundreds of court cases on redistricting.\footnote{See for instance
\emph{Louisiana House of Reps.~v Ashcroft}, 539 U.S.~461 (2003); 
\emph{Martinez v Bush}, 234 F.~Supp.~2d 1275 (S.D.~Fla.~2002); 
\emph{Perez v Perry}, 835 F.~Supp.~2d 209, 211 (W.D.~Tex.~2011); 
\emph{Vesilind v Virginia State Board of Elections}, 15 F.~Supp.~3d 657, 664 (E.D.~Va.~2014); 
\emph{Page v Virginia State Board of Elections}, 15 F.~Supp.~3d 657 (E.D.~Va.~June 5, 2015); 
\emph{Sanders v Dooly County}, 245 F.~3d 1289 (11th Cir.~2001); 
\emph{Session v Perry}, 298 F.~Supp.~2d 451 (E.D.~Tex.~Jan 6, 2004); 
\emph{Garza v County of Los Angeles}, Cal., 756 F.~Supp.~1298 (C.D.~Cal.~1991); 
\emph{Harris v McCrory}, 159 F.~Supp.~3d 600, 611 (M.D.N.C.~2016); 
\emph{Johnson v Miller}, 922 F.~Supp.~1552 (S.D.~Ga.~1995); 
\emph{Cromartie v Hunt}, 526 U.S.~541 (1999); 
\emph{Moon v Meadows}, 952 F.~Supp.~1141 (E.D.~Va.~1997); 
and many more.}

A cosmetic variant of the Polsby-Popper score, which in fact predates Polsby-Popper in the literature, is the \emph{Schwartzberg score}. This was 
originally
defined as the ratio of the perimeter of a district to the perimeter (circumference) of the circle having the same area---clearly echoing the Polsby-Popper construction---so that  lower Schwartzberg scores are deemed preferable to higher ones \cite{schwartzberg}.  
This is expressed by
$${\sf Schw}(\Omega):=\frac{\perim(\Omega)}{\sqrt{4\pi\cdot \area(\Omega)}} = \PP(\Omega)^{-1/2}.$$
Since {\sf Schw}  is simply a power of \PP, 
Schwartzberg and Polsby-Popper assessments must rank districts from best to worst in precisely the same way.\footnote{This is because for positive values of $x$ and $y$, we have $x>y \iff x^{-1/2}<y^{-1/2}$. Therefore a higher (and thus better) $\PP$ score corresponds to a lower (and thus better)  {\sf Schw} score.} 
But because Joseph Schwartzberg worried that 
there was no way (with 1966 technology) to accurately measure perimeters of districts, 
he also proposed a notion of {\em gross perimeter} using a simplified boundary to make the problem more tractable \cite{schwartzberg}. Because it follows the article literally, Maptitude redistricting software uses a different definition of perimeter in the computation of a Schwartzberg score than in the computation of a Polsby-Popper score, which of course  breaks the scores' monotonic relationship and once again highlights the power of software to create and stabilize definitions.

\subsection{The landscape of compactness metrics}

\begin{figure}[htbp]
\centering
\includegraphics[width=5in]{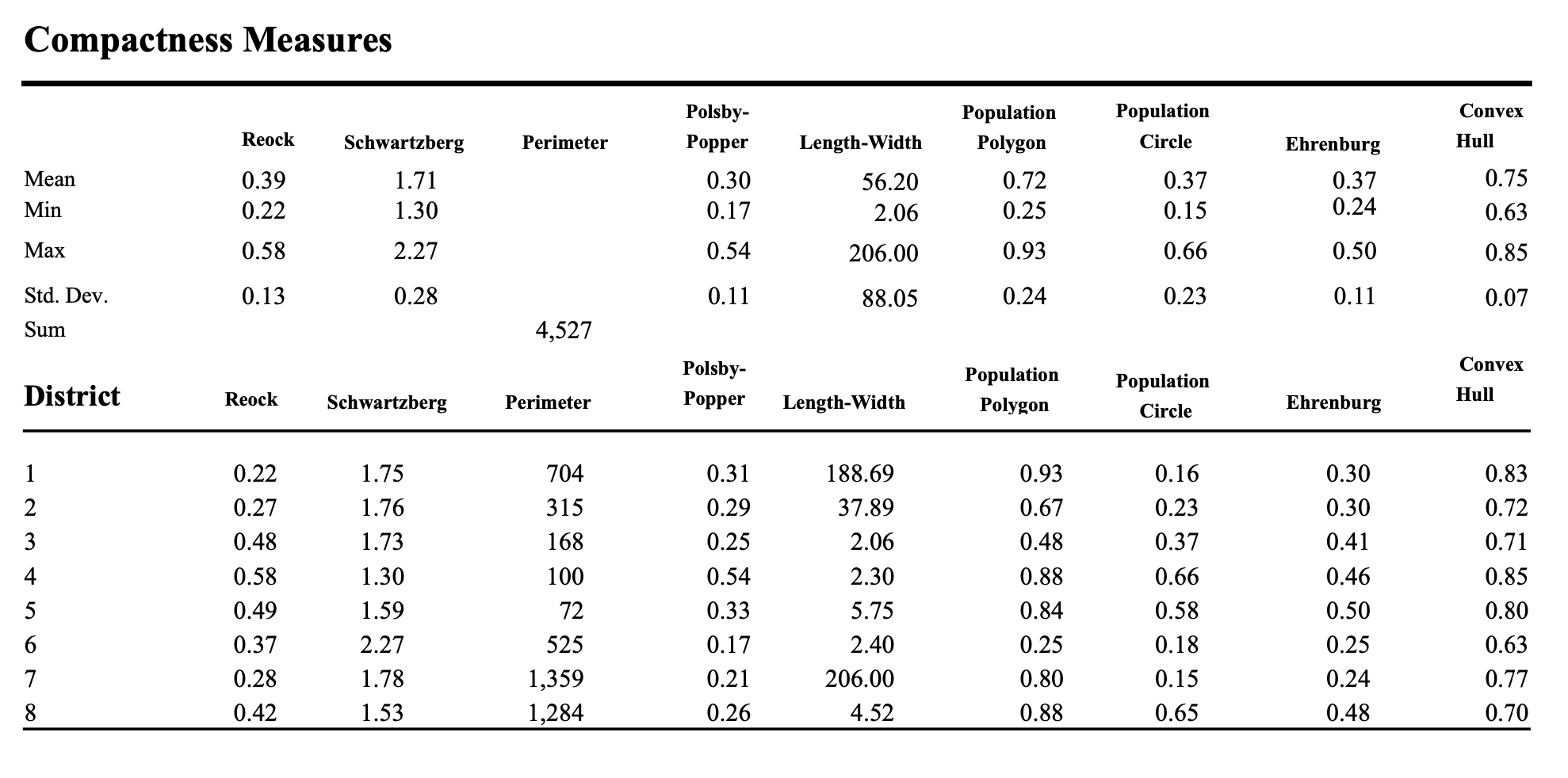}
\caption{Page from a report generated by Maptitude for Redistricting, reporting  Maptitude's standard compactness scores for a Minnesota districting plan.}\label{fig:maptitude}
\end{figure}

Despite the fact that experts frequently cite Polsby-Popper scores, there is no consensus on how these scores should be used 
when determining the validity of a districting plan.
To make matters more confusing, legal contexts often call for the reporting of more than one type of compactness
score.
Consider the 2018 litigation-driven 
congressional redistricting in Pennsylvania.  In the court orders of January 22 and 26 that year,
it was declared that ``[A]ny redistricting plan the parties or intervenors choose to submit to the Court for its consideration shall include $\ldots$ [a] report detailing the compactness of the districts according to each of the following measures: Reock; Schwartzberg; Polsby-Popper; Population Polygon; and Minimum Convex Polygon.''\footnote{\emph{Turzai v League of Women Voters of Pennsylvania}, 17A909 (2018).}   The last three of these metrics are defined as follows.
\begin{itemize}
\item {\bf Reock}:  the area of a district divided by the area of its smallest circumscribing circle \cite{reock};
\item {\bf Population Polygon}: the population of a district divided by the population contained in its convex hull;\footnote{A \emph{convex body} is a region that contains the entire line segment between any two of its points. 
The \emph{convex hull} of a region is the smallest convex body containing 
the entire region.  This is sometimes picturesquely
referred to as the ``rubber-band enclosure.''} and
\item {\bf Minimum Convex Polygon} (also known as the Convex Hull score):  
the area of a district divided by the area of its convex hull.
\end{itemize}
The court orders do not specify whether any of these assessments might be more important than the others, nor how two plans are to be compared.  
If two plans were being evaluated in terms of their Reock scores, each plan had 18 values to consider, and it is not obvious how to say that one suite of scores is better than the other (see Figure~\ref{fig:maptitude} for 8-district Minnesota comparisons).
Averaging over the districts to make two plans directly comparable may give some insight, but it fails to distinguish a plan where all districts are moderate from another where some districts score extremely badly 
while others  score favorably.  

Each of these metrics, including Polsby-Popper and Schwartzberg, requires rendering a district as a domain on a map of the state. This domain is bounded by a  contour, and then classical (i.e., Euclidean) plane geometry is invoked to make some sort of 
computation.  Population Polygon stands out by taking population location into account, but it still relies on the contour in a fundamental way.  In fact, the 1990 survey \cite{NGCH} identifies 24 compactness metrics, and every one of them depends on a planar embedding, including the three that are designated as population-based.

A focus on contours is not fundamental to the geography or the geometry
of partitions. Since the early twentieth century, geometry has flourished in a {\em discrete}, \emph{combinatorial} setting. The objects in this framework (such as {\em graphs}, {\em groups}, and {\em complexes}) 
are made up of individual units that one can enumerate and identify, rather than the smooth curves of classical geometry.
Because of the small menu of important geographical units in this application---principally counties, precincts, and blocks---discrete geometry gives a useful toolkit for the geography of redistricting, as we will see below in \S\ref{sec:discrete}.  

\section{Problems with contour-based scores}\label{sec:issues}

To fully appreciate the benefits of a discrete perspective on this problem, we set the stage by itemizing some drawbacks to contour-based compactness. 
Contour-based scores face four primary issues common to the category:  
trouble with {\bf physical geography}, {\bf resolution instability}, {\bf coordinate dependence}, and {\bf areal emphasis}.  
We discuss these below, 
together with examples that are mainly drawn from the the 113th Congress (2013) vintage using the repository of data and code found in \cite{DiscCpct}.
The goal is to highlight that the definitions are shakier than they seem, and that the scores often align poorly with commonly understood best practices in redistricting.

As already discussed, the classic scores often face challenges of aggregation and comparison.\footnote{A notable variant to these district-level scores is simply to report the total perimeter involved in a districting plan.
For example,  the state constitutions of Iowa and Colorado and at least one expert 
report (\emph{Puerto Rican Legal Defense and Education Fund v Gantt}, 796 F.~Supp.~677 (E.D.N.Y.~1992))
compare the total area of the jurisdiction (which is constant across alternative/contending districting plans for that jurisdiction) to the sum of all district perimeters.   This handles the aggregation problem but not the other drawbacks of contours.}  
For instance, the five scores cited in the Pennsylvania litigation are all valued between 0 and 1, inviting comparisons across place, scale, and time.  
Though any careful practitioner would caveat the use of averages and would avoid giving the impression that a district score of $.209$ in one context could be meaningfully compared to a district score of $.205$ in another, courts often reach for simple summary statistics and direct comparison.  We include some naive numerical comparisons below, such as the rank among the 435 Congressional districts, to highlight the risks of uncritical quantification.

\coastline

\begin{figure}[htbp]
\centerline{\includegraphics[height=2in]{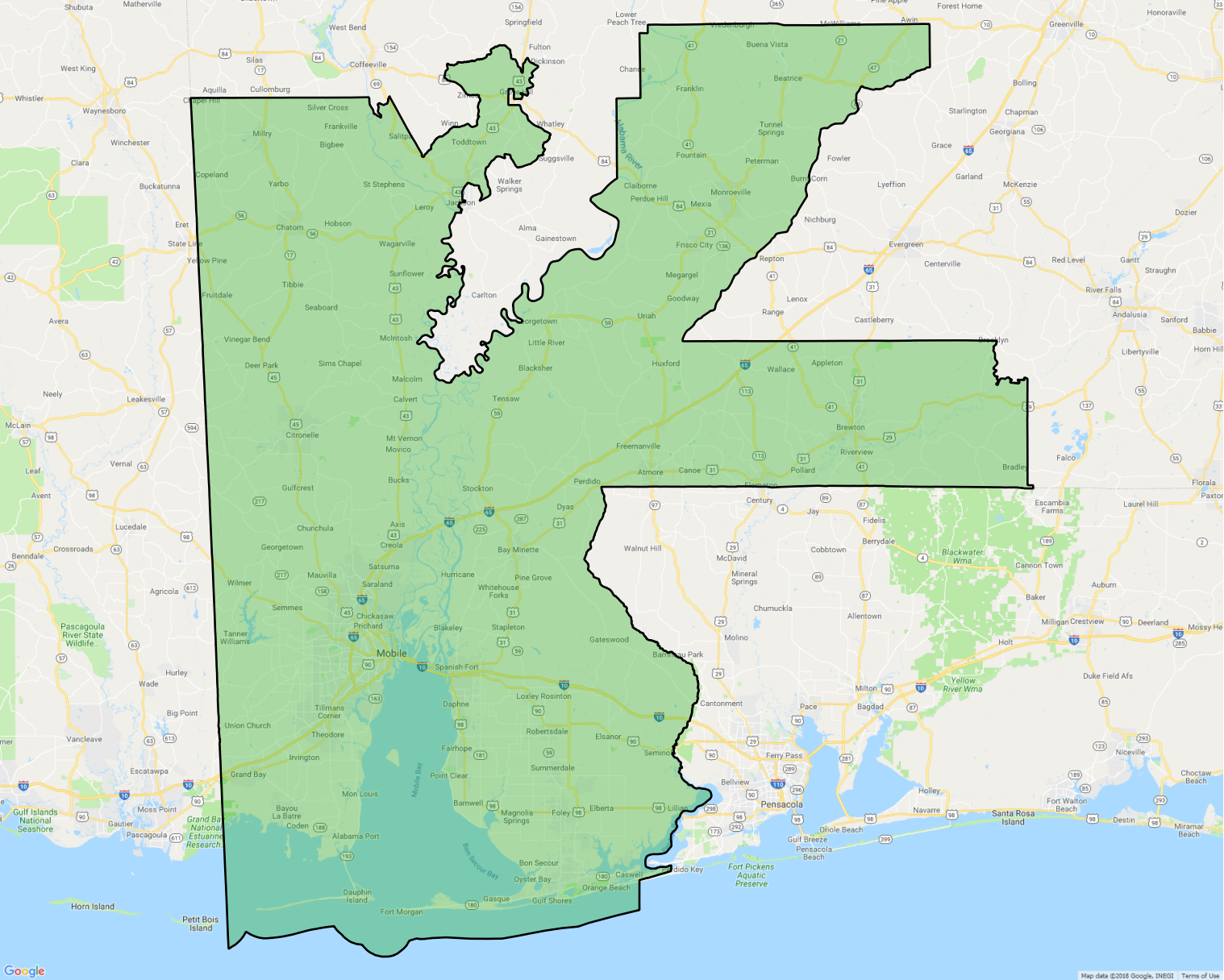} \qquad
\includegraphics[height=2in]{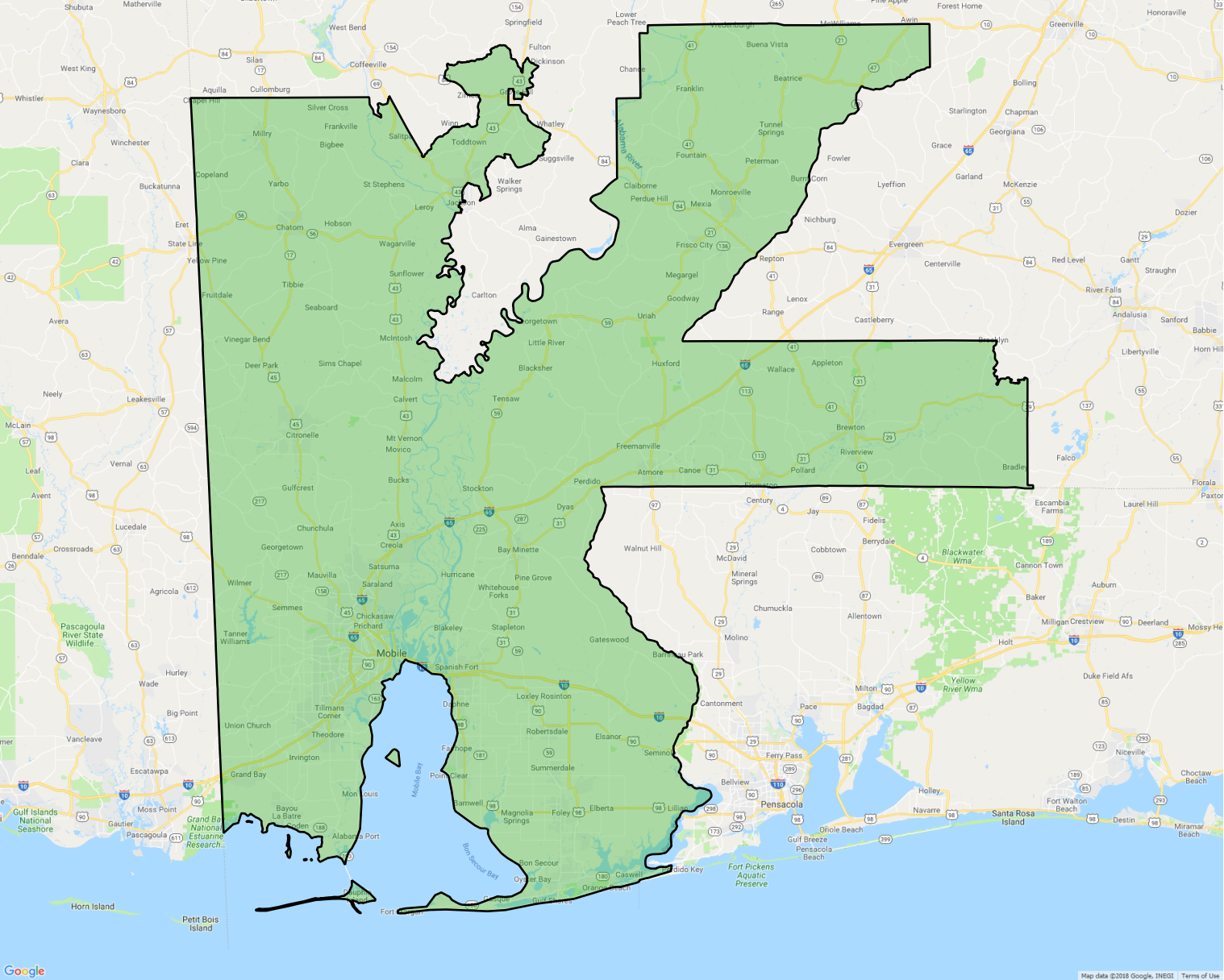}}
\caption{Alabama's 1st district has boundary partly defined by  the Gulf of Mexico and the Tombigbee and Alabama rivers.  The TIGER/Line (left) and Cartographic 500K (right) maps are shown here, illustrating that
the Polsby-Popper quantification of compactness leaves the modeler caught in an unpleasant choice between
a map subject to coastline effects (Issue A) or to arbitrary inclusion of unpopulated areas (Issue D).}\label{fig:AL-1}
\end{figure}

The perimeter of a contour-based region cannot account for a pertinent geographical feature like a coastline or an irregular state boundary that explains a portion of a district's border. The districting plan might incur a steep penalty for having an erratic perimeter, even though that border was not chosen through any questionable or manipulative process. 
For example, Alabama's 1st Congressional District from 2013 is partly bounded by the Gulf of Mexico to the south, and the Tombigbee and Alabama Rivers to the north. As depicted in Figure~\ref{fig:AL-1}, this creates sections of eccentric natural boundary.  
Some shapefiles mitigate the effects from Gulf boundary by adding a buffer extending into the water---but of course
nothing similar can be done for the river boundary.
Accordingly, AL-1 has a fairly low Polsby-Popper score of approximately $.162$, ranking 318th out of 435 districts in the TIGER/Line Shapefiles
(shown at left in the figure), but scores significantly worse ($.111$, ranking 367th) in the Cartographic maps (shown at right).\footnote{The reader can find code, data, and documentation 
for area and perimeter statistics at \cite{DiscCpct}.}
There is no standard on whether to include water when reporting compactness scores.\footnote{For 
instance, some districting plans filed with the court in Pennsylvania's 2018 redistricting  included portions of Lake Erie in the northwest of the state, while others did not.}

\begin{figure}[htbp]
\centerline{\includegraphics[height=1.2in]{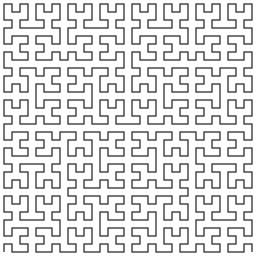} }
\caption{A Hilbert space-filling curve in an intermediate stage of construction.  The winding becomes progressively more complicated, finally converging on a curve of infinite length that fits in a finite area.}
\label{fig:spacefillingcurve}
\end{figure}

Consideration of the coastline issue leads naturally to a related worry about stability of scores under changes
in resolution.
A coastline border is irregular and, in a sense, unmeasurable. This is the well-known ``coastline paradox'' sometimes attributed to Benoit Mandelbrot:  the length of the coast of Great Britain depends on the size of one's ruler  \cite{mandelbrot}. In this way, the quantities $\area(\Omega)$ and $\perim(\Omega)$ depend on the scale of precision used when mapping the region, and can change significantly at different levels of zoom. Finer wiggles in the boundary can expand the perimeter with no limit. Indeed, arbitrarily long perimeter can exist within a fixed finite area, as shown in Figure~\ref{fig:spacefillingcurve}.

The penalty on district borders that follow natural features is contrary to good districting practices, particularly with respect to the goal of easily described district lines.  Natural borders like rivers, which often coincide with town, county, and state lines, are clearly far preferable to lines or arcs that do not correspond to visible or marked features.\footnote{Census geography is built with this in mind, as for instance with census blocks, which are  ``formed
by streets, roads, railroads, streams and other bodies of water, other visible physical and cultural features, and the legal boundaries shown on Census Bureau maps'' \cite{glossary}.}

Next, we consider what the cartography and GIS literature refers to {\em simplification} and {\em generalization}:  varying the number of points and features that define a curve.  In computing, this would be considered a matter of {\em resolution}. The geography and computer science communities share a concern with the effects of scale on precision.

\stability

Consider the
Census Bureau Cartographic Boundary files, which are available in three scales:
\begin{center}
500K (1:500,000), \quad 5M (1:5,000,000), \quad 20M (1:20,000,000).\end{center}
One would expect some variation in the perimeters and areas of districts because the 20M files are greatly simplified. Indeed, the Census Bureau itself flags this issue,
warning that 
 ``These boundary files are specifically designed for small scale thematic mapping $\ldots$
These files should not be used for $\ldots$ geographic analysis including area or perimeter calculation'' 
\cite{cartographic}.
Nonetheless, we use those maps here as an extreme illustration of an issue that will be present
whenever map resolution can vary:  not only are area and perimeter themselves altered, but those 
changes are compounded by the way Polsby-Popper is calculated.  Perimeter is typically more sensitive
to resolution change, and because it is squared, the Polsby-Popper score may drop precipitously at higher resolutions.

For example, California's 53rd Congressional District, an unremarkable-looking district located inland in San Diego County before its elimination in the last reapportionment, saw an 81\% jump in perimeter when going from the 20M scale to the 500K scale. In the same transition, the district's area increases by less than 9\%. 
This has a major effect on the district's relative ranking of \PP score
among the 435 congressional districts:  from 
ranking 61st at the coarsest zoom, it drops to 191st at the intermediate resolution, and then to 292nd at the finest zoom.  That means
that the district's assessed shape quality goes from being in the best third, to the middle third, to the worst third, inviting  completely different qualitative 
assessments.
On average, when comparing data between the 20M scale and the 500K scale, congressional district perimeters increase by about 23\%, while district areas increase by 0.2\%. 
Clearly both statistics are sensitive to resolution, and perimeter is markedly more so.\footnote{In \cite{BS}, Barnes and Solomon  explore the resolution sensitivity, bypassing Census cartographic maps by varying map resolution along a spectrum and using a sophisticated geometric toolkit.}
Census Bureau TIGER/Line files are updated with slight modifications every year; even state boundaries are regularly adjusted, sometimes with adjustments on the order of inches from one year to the next, reflecting what the Bureau regards as improved accuracy.
Redistricting analysts would need to use maps not only from the same source, but also from precisely the 
same vintage, in order to expect consonant 
results.

\coord

\begin{figure}[htbp]
\centerline{\includegraphics[height=1.3in]{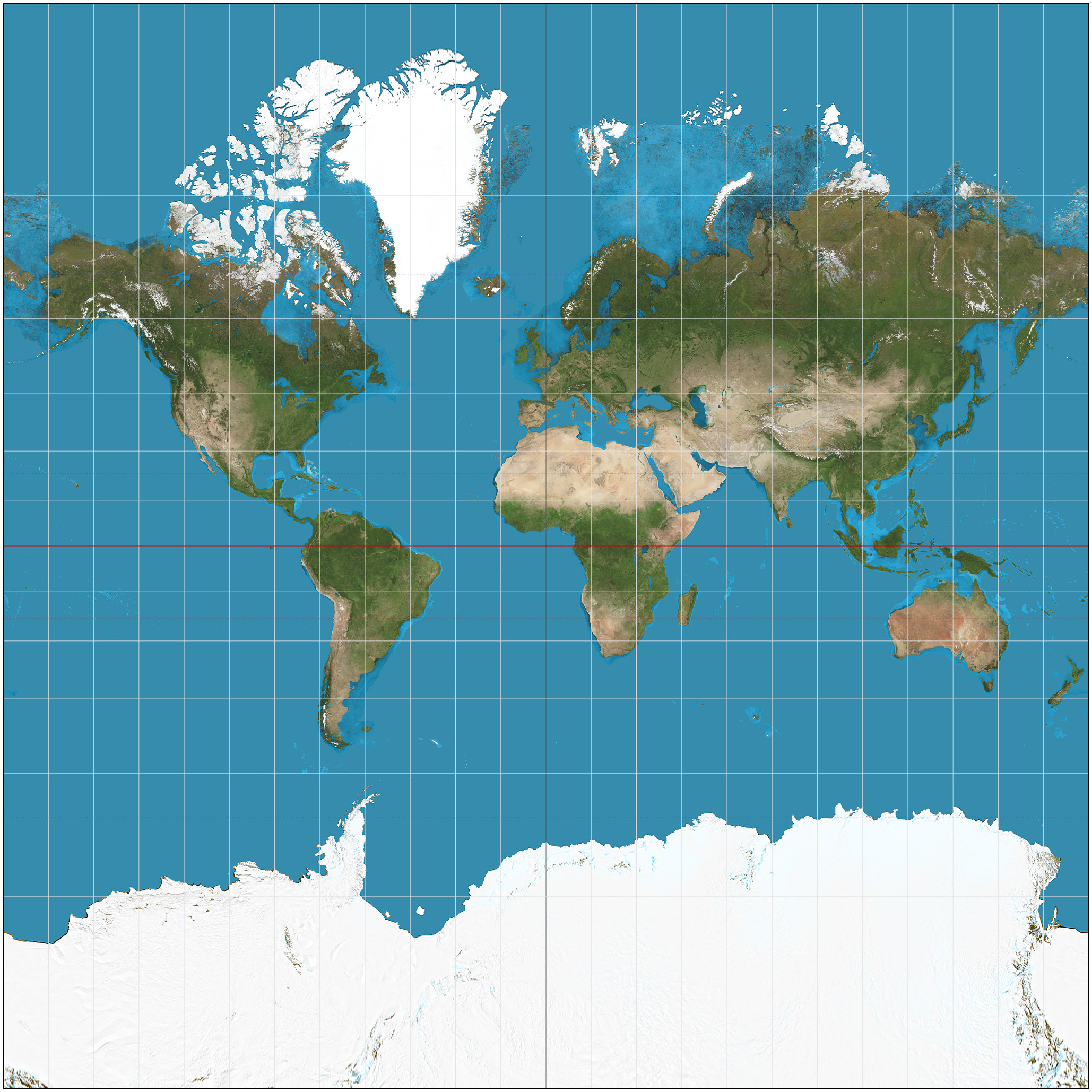}\ \
\includegraphics[height=1.3in]{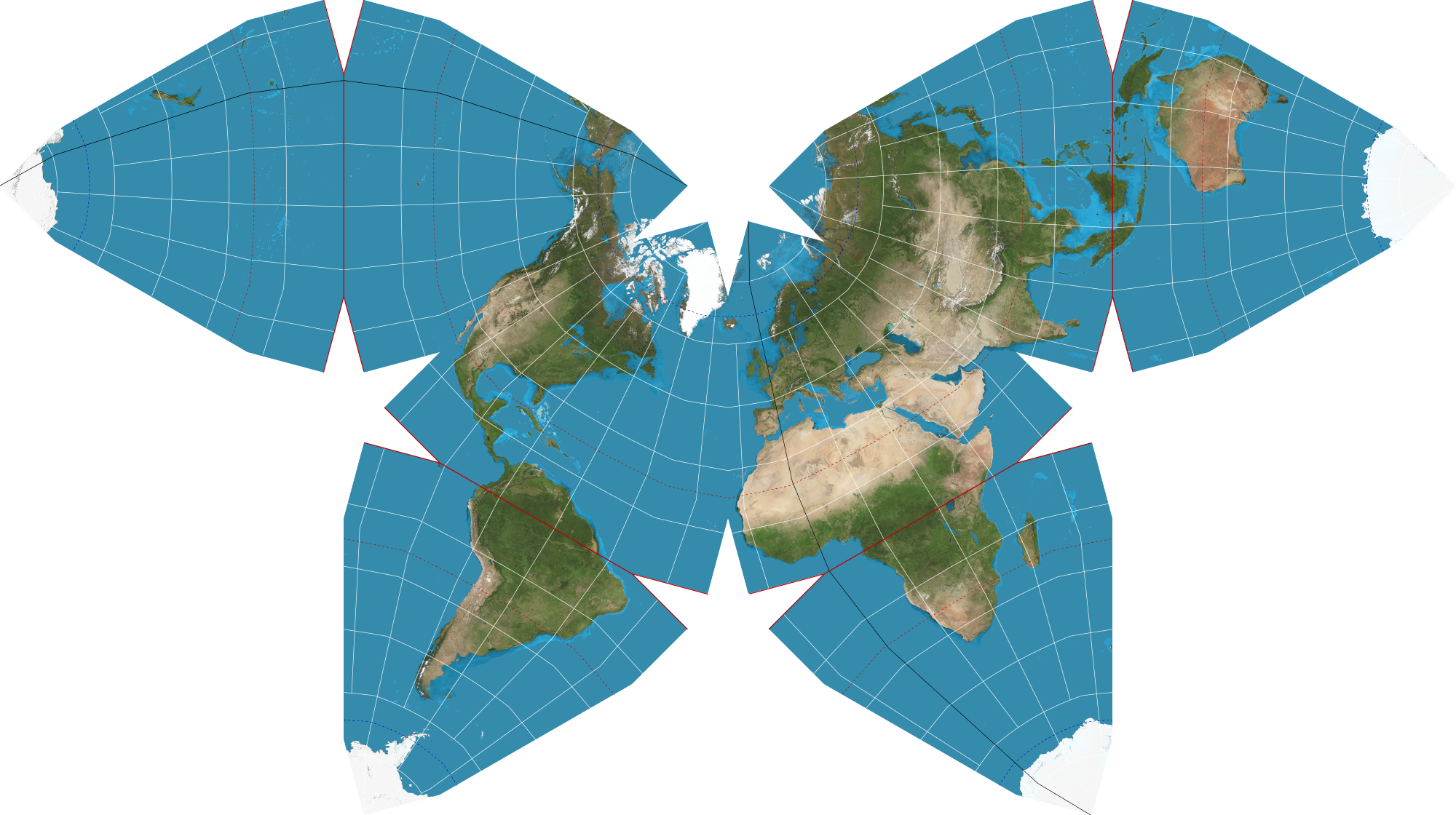} 
\includegraphics[height=1.3in]{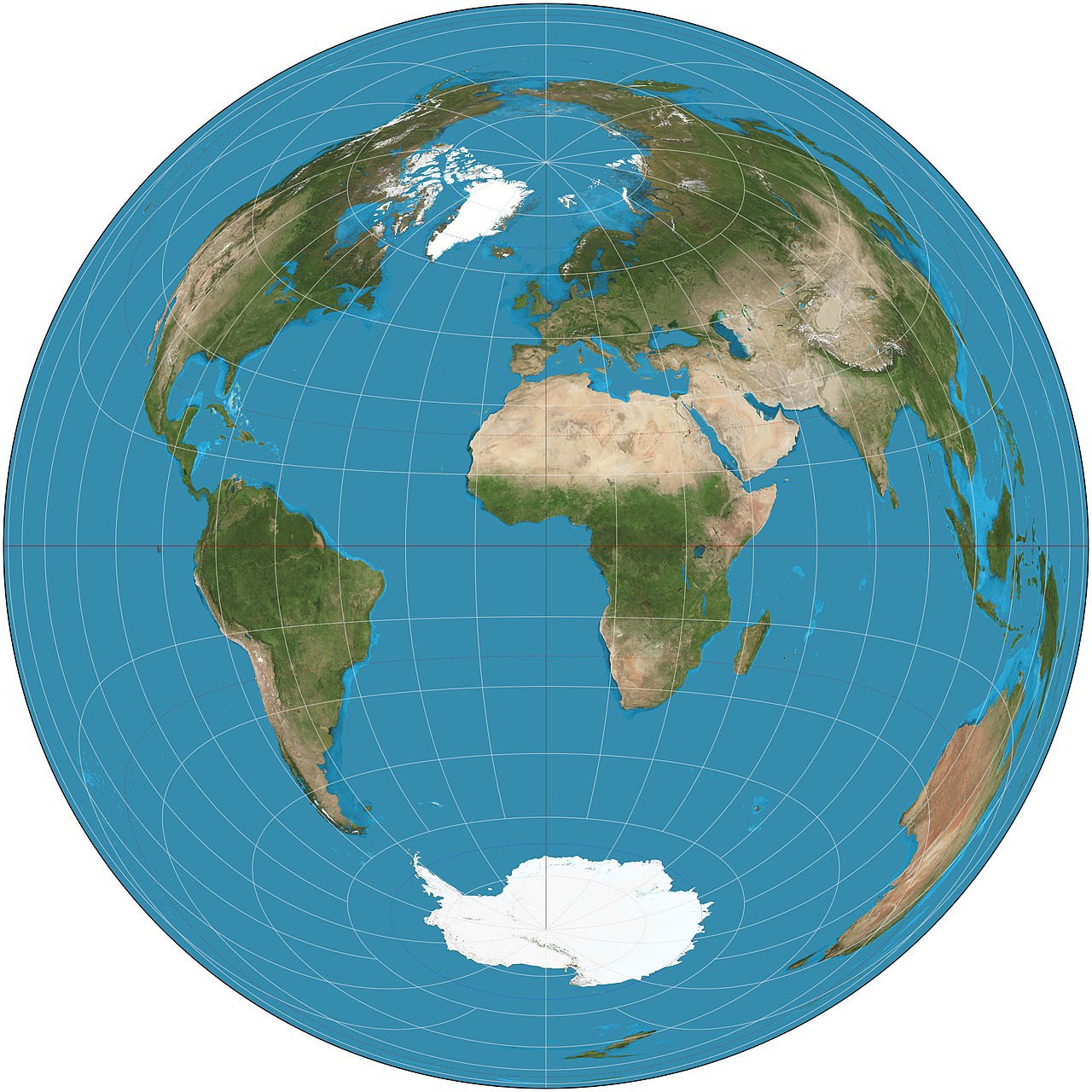}}
\caption{The first map projection preserves angle measurements from the sphere,  the last preserves area, and the middle map 
is an attempted compromise between the two, sacrificing some accuracy in each \cite{kunimune,strebe}.}\label{fig:projections}
\end{figure}

 Geographers have a well-established apparatus for specifying {\em projections} from sphere to plane: the Earth is round, roughly speaking, but most maps are flat.    It is widely known to be impossible
to choose a map projection that faithfully preserves both shape and  area of regions on the sphere. 
From a mathematical perspective, the proof of this fact is simple to sketch:  any smooth map that preserves area and angles must be a {\em local isometry} (i.e., it preserves distances at small scale), and so must preserve total curvature.  Spheres have positive total curvature while planes have zero total curvature, making such a map impossible. Three example projections are shown in Figure~\ref{fig:projections}.

A recent paper by Bar Natan, Najt, and Schutzmann \cite{GerryJumble} beautifully proves what one might call an impossibility theorem for consistent map projection from a mathematical point of view.  The authors show that any possible map projection will reverse the order of some pair of districts with respect to Polsby-Popper scores, Reock scores, and convex hull scores.  That is, if you specify any map projection, you can always find a pair of regions so that one scores better on the sphere, but the other scores better once you have projected to the plane.  

The Reock score is not as badly plagued by coastline and resolution issues as Polsby-Popper because it depends on area and on the circumscribing circle, both of which are fairly robust to small boundary perturbations.  On the other hand, Reock is flagged as having especially strong coordinate dependence in \cite{GerryJumble}.
Hachadoorian et al.~\cite{hachadoorian} also single out Reock scores for their extreme projection-dependence.
For instance, out of 18 districts that they selected for comparison, 8 had changes of 24\% or more in 
their Reock scores among three projected coordinate systems
(locally parametrized Albers equal-area; World Mercator; and plate carr\'ee lat-long).  Worse,
the changes were in unpredictable directions, with some scores increasing and some decreasing over a given
shift in map projection.  Since there is certainly no standard map projection in the redistricting use case---for instance, Maptitude seems to use a locally parametrized CRS while the popular Dave's Redistricting App uses lat-long coordinates---this dependence undermines the meaningfulness of contour-based scores.
   
Even the Population Polygon score, which sounds promising because it is population-based, suffers from coordinate dependence. The projection of the district does  not impact the population of the district, of course, 
but it heavily affects the form of the convex hull, and therefore the population enclosed by it.

\emptyspace

Districts are to be equalized by population, not by acreage, and districts are intended to specify a division of voters.  Consider an unpopulated geographical region---a designated wilderness area, say---with different districts to its north and south. The assignment of any part of this unpopulated region to the northern district has no influence on voting, and should not have a great impact on the districts' quality. However, the choice of how to allocate this unpopulated region between the two districts will have a marked influence on their perimeters and areas.

\begin{figure}[htbp] 
\includegraphics[height=1.75in]{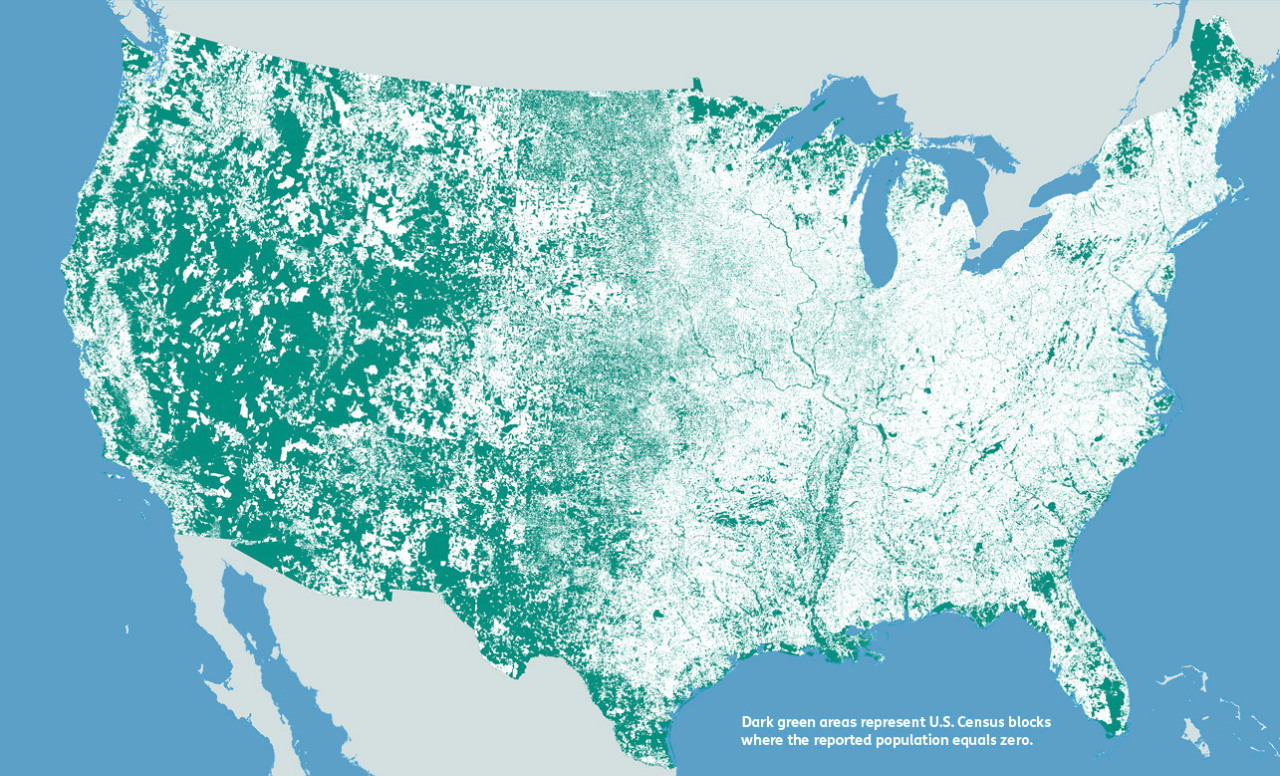} \hfill
\includegraphics[height=1.75in]{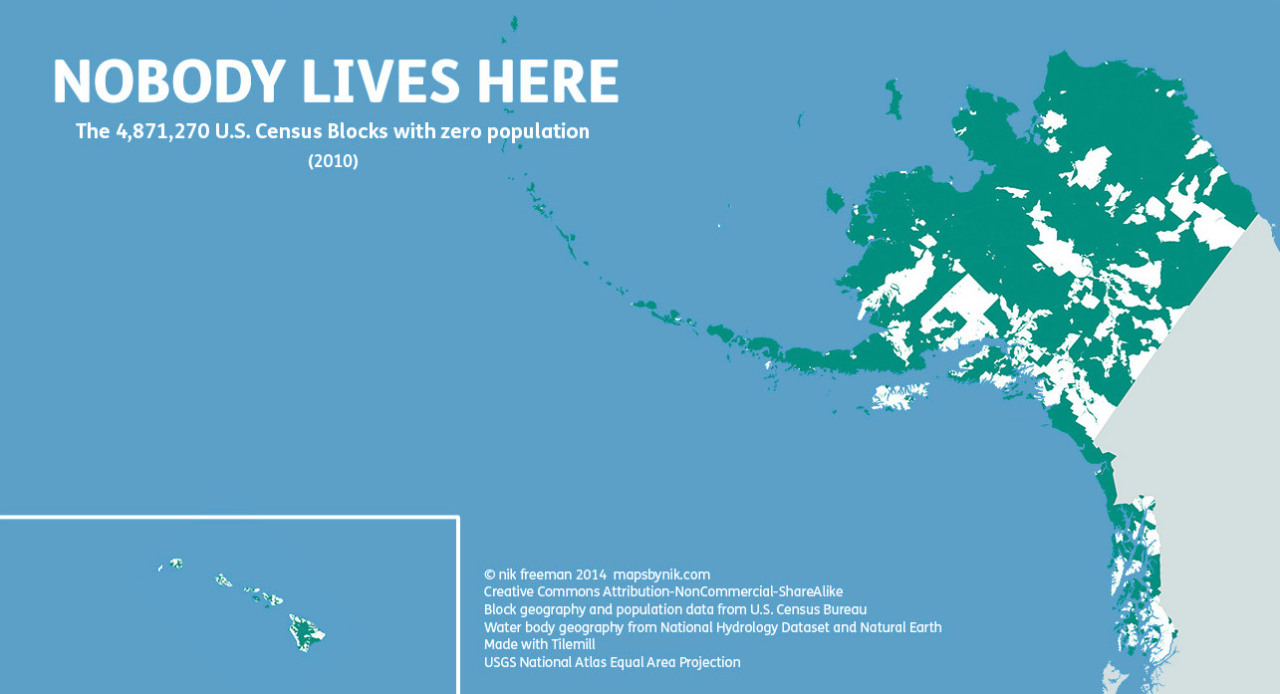}
\caption{Unpopulated census blocks, excluding the Great Lakes, 
are depicted in dark green in these maps by Nik Freeman \cite{mapsbynik}.}\label{fig:empty blocks}
\end{figure}

One major source of unpopulated surface area is water.
To see the impact that this can have on shapes, consider that by Census measurements,  fifteen states are at least 10\% water by area, with Michigan topping the list at 41.5\%.
Overall, water makes up 7\% of the United States census geography \cite{census-area}.  
More broadly, in the 2010 Census, nearly 45\% of  blocks had zero reported population;
these blocks that were unpopulated after the 2010 census are depicted in Figure~\ref{fig:empty blocks}.
Compactness can be wildly skewed by  assignments of officially unpopulated area to districts.  

Cartographers are of course attuned to this issue, so for instance  conventional (areal) representations are often complemented with cartograms that are resized by population.  But it is important to remember that attempts to mitigate areal emphasis in district compactness scores are going to butt up against the ``eyeball test.''
This will prevent us from leaving unpopulated areas completely out of any scoring system; since they have a major impact on the optics of a district, they will have to figure into any metric that hopes to pass muster as a compactness score.

\section{Discrete geometry and discrete compactness}\label{sec:discrete}

This section contains further motivation for discrete metrics, as well as formal definitions and examples.  We present two specific discrete metrics in \S\ref{sec:cut-edges} and \S\ref{sec:spanning-trees}.

\subsection{Discreteness and graphs}
The mathematical term {\em discrete} refers to a set whose elements are distinguishably isolated from each 
other.\footnote{The technical definition is as follows: given a topological space $X$, a subset $S \subseteq X$ is \emph{discrete} if for each element $x \in X$, there exists an open neighborhood $U(x)$ containing $x$ and no other element of $S$. For instance, the integers $\{\dots,-1,0,1,2,\dots\}$ are discrete as a subset of the number line because a small enough interval around one integer will separate it from the others.  By contrast, consider the set of the rational numbers $p/q$ on the number line:  there is no way to isolate a single one of these points from all others, no matter how closely one zooms in.}
Any set  with only finitely many elements (in a space with a notion of distance) 
is necessarily discrete. 
The relationship between redistricting and standardized Census geography means that redistricting is necessarily a finite---although perhaps gigantic---problem, and therefore discrete.

For redistricting purposes, it is  important to know which geographical units 
are adjacent to which other ones,
in order to make sense of district contiguity.
The mathematical abstraction for recording a discrete set of elements and the adjacencies among them is called a 
{\em graph} or {\em network}.  
Here we will introduce only as much terminology as is needed for this discussion, and refer the reader to numerous graph theory texts (for example, \cite{diestel}) for more information.

Formally, a (simple undirected) \emph{graph} 
$G=(V,E)$ consists of a \emph{vertex set} $V$ and an \emph{edge set} $E$. Each edge is an unordered pair of distinct vertices, said to be  \emph{adjacent} to each other. See the second and third drawings in Figure~\ref{fig:dual} 
for examples of such 
graphs.  
The {\em degree} of a vertex is the number of edges that are incident to it.
Our graphs will be endowed with a (vertex) \emph{weight function}, $w : V \rightarrow \mathbb{R}_{\ge 0}$, associating nonnegative values to the vertices. 

\subsection{Creating geography dual graphs}
We now have notation and terminology needed to build a population-weighted graph that is ``dual'' to a set of geographic units.  
This graph will contain one vertex for each unit.  We put an edge between two vertices when the corresponding  units share part of their boundary; note that a model requires a choice between so-called 
{\em rook adjacency} or {\em queen adjacency} to build the graph, where the names are drawn from the movement of the 
corresponding chess pieces (see Figure~\ref{fig:dual}).
This is a standard construction called the {\em dual graph} of a planar tiling.

\begin{figure}[htbp]
\begin{center}\begin{tikzpicture} 
\draw [fill=blue!50]  (0,1) rectangle (1,2); 
\draw [fill=green!30!blue!20]  (1,1) rectangle  (2,2); 
\draw [fill=red!50] (0,0) rectangle (1,1) ;
\draw [fill=orange!50] (1,0) rectangle  (2,1) ;
\draw [fill=yellow] (2,0) rectangle (3,2);

\begin{scope}[xshift=5cm] 
\draw (1/2,1/2) rectangle (3/2,3/2) (3/2,3/2)--(5/2,1)--(3/2,1/2);
\draw [fill=blue!50] (1/2,3/2) circle (1/5);
\draw [fill=green!30!blue!20] (3/2,3/2) circle (1/5);
\draw [fill=red!50] (1/2,1/2) circle (1/5);
\draw [fill=orange!50] (3/2,1/2) circle (1/5);
\draw [fill=yellow] (5/2,1) circle (1/5);
\node at (3/2,0) [black] {dual graph};
\node at (3/2,0) [below=5pt,black] {(rook)};
\end{scope}

\begin{scope}[xshift=10cm] 
\draw (1/2,1/2) rectangle (3/2,3/2) (3/2,3/2)--(5/2,1)--(3/2,1/2);
\draw (1/2,1/2) -- (3/2,3/2) (1/2,3/2)--(3/2,1/2);
\draw [fill=blue!50] (1/2,3/2) circle (1/5);
\draw [fill=green!30!blue!20] (3/2,3/2) circle (1/5);
\draw [fill=red!50] (1/2,1/2) circle (1/5);
\draw [fill=orange!50] (3/2,1/2) circle (1/5);
\draw [fill=yellow] (5/2,1) circle (1/5);
\node at (3/2,0) [black] {dual graph};
\node at (3/2,0) [below=5pt,black] {(queen)};
\end{scope}
\end{tikzpicture}
\end{center}
\caption{On the left is a tiling of a rectangular region into five units. The middle and right-hand figures represent dual graphs of this tiling, using rook and queen adjacency, respectively---the queen style allows corner connections, considering two tiles meeting at a point to be adjacent. 
Some states specify which form of contiguity is permitted in their districts.}\label{fig:dual}
\end{figure}

\begin{figure}[htbp]
\centering
\includegraphics[width=2.3in]{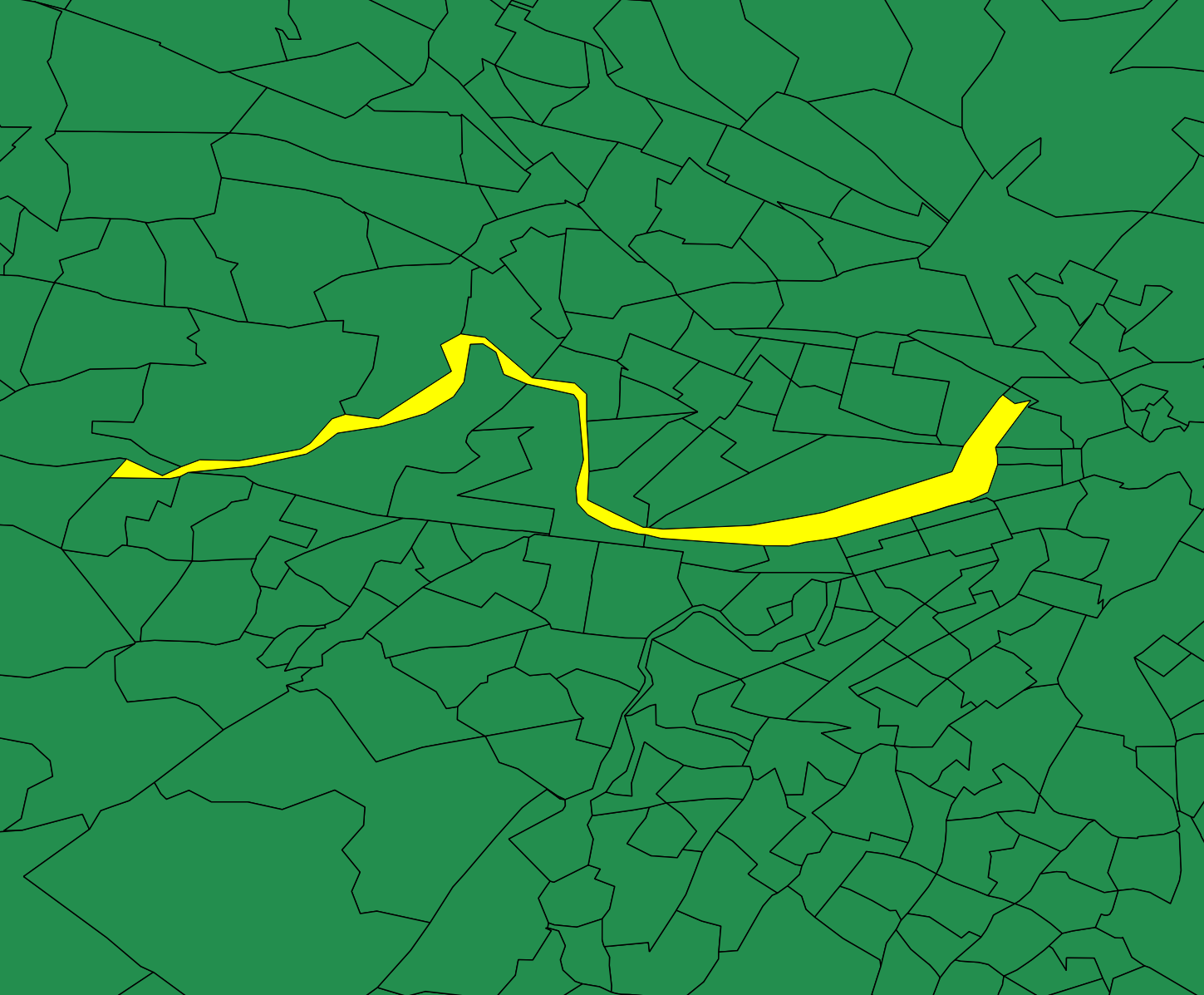}
\caption{Census tracts in the Boston area, with tract 981501 highlighted.}\label{fig:tracts} 
\end{figure}

The census data comes with population counts on the geographical units, which we will record with a
weight function on the vertices.  For instance, the segment of the Charles River near Allston, MA is a census tract designated by the GEOID 981501 (see Figure~\ref{fig:tracts}), and the 2010 Census says it has population 12. So in a dual graph on tracts, we could write $w(981501)=12$ to indicate that population count as a weight on the vertex labeled 981501.  This tract is rook-adjacent to twenty-one other tracts, so the vertex labeled 981501 would have degree $21$ in that graph. It is not unusual for water units, less subdivided than land, to have high degree in this way.

We can define the population for a set of vertices by summing:  if $A\subseteq V$, we write
$w(A):=\sum_{v\in A} w(v)$  for the total population of all the units that make up $A$.  So for instance if 
$S$ is the set of vertices corresponding to census tracts in Suffolk County, MA, within the graph of Massachusetts tracts, then $w(S)$ is the population of the county.

\subsection{Plans as partitions}\label{sec:representing}

\begin{figure}[htbp]
\centering
\includegraphics[height=2in]{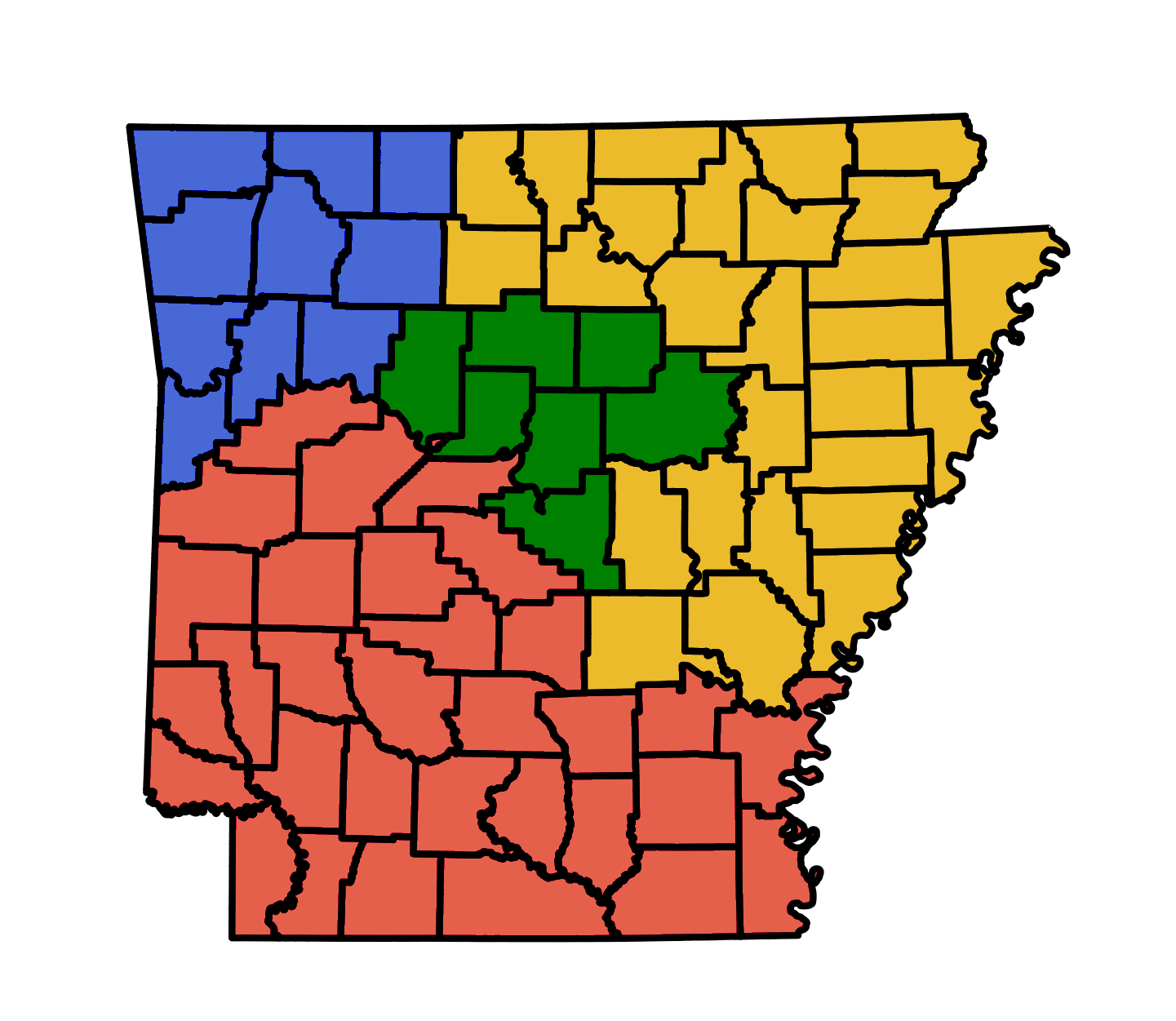} \qquad \includegraphics[height=2in,width=2.2in]{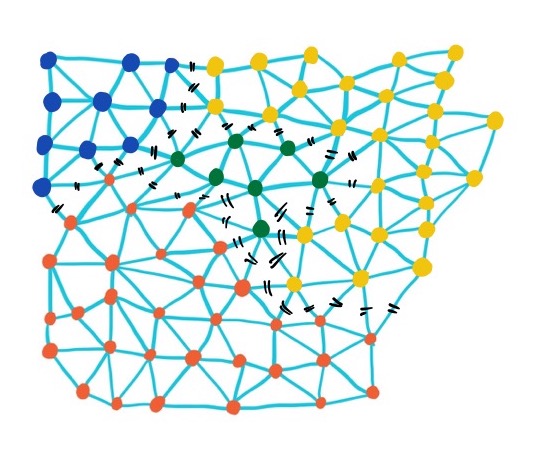}
\caption{This figure shows two representations of  a partition of the counties of Arkansas into four districts ($k=4$).  In this plan, each district is within about 4\% of ideal population, so in particular this could be considered a balanced partition at $\epsilon=.05$.
The county dual graph shown on the right has 75 vertices (one for each county) and 192 edges (one for each pair of adjacent counties).
This plan has 36 marked edges between 
pairs of counties that are adjacent but that belong to different districts (Blue/Yellow: 3, Blue/Green: 2, Blue/Red 4, Red/Green: 8, Red/Yellow 6, Yellow/Green: 13).
That means that 36 of the 192 adjacencies have to be severed in order to separate the state into districts in this pattern; below we define this as the cut edge count, or cut score, of the plan.  
Figure reprinted from \cite{gerrybook} with permission.
}\label{fig:arkansas}
\end{figure}

By a {\em partition} of a graph $G$, we mean a decomposition into mutually disjoint subgraphs $P_1,\ldots, P_k$ 
that, between them, cover all of the vertices of the graph.\footnote{In mathematical notation, a vertex partition is given by $V=V_1 \cup \dots \cup V_k$ where 
$V_i\cap V_j=\emptyset$ for all $i\neq j$. 
For each $V_i$, the {\em induced graph} $P_i$ has vertex set $V_i$ and includes all edges between those vertices that were present in the original graph $G$.  
}
Our notation for a partition will be $P=(P_1,\dots, P_k)$.  If each subgraph $P_i$ is connected, then we call each $P_i$ a  {\em district} and we call $P$ a {\em districting plan} (or simply a {\em plan}).  
For instance, in Figure~\ref{fig:arkansas}, we could take $P_1$ to be the subgraph on the yellow vertices, $P_2$ the subgraph on the green vertices, and so on.  
Below, we will assume that a graph $G$ has been fixed, so that we can talk about its partitions without referring back to $G$ in the notation.

Now we can restrict districting plans with $k$ districts to those that balance the census population to within a tolerance 
$\epsilon\ge 0$.  
That is, we require
$$(1-\epsilon) \frac{w(G)}{k} \le w(P_i) \le (1+\epsilon) \frac{w(G)}{k}$$
for all districts $P_i$ in the plan.  If $\epsilon=.05$, for example, this amounts to requiring that each district be within 5\% of ideal district population.

\subsection{Cut edges}\label{sec:cut-edges}

The first discrete shape measure that we present is the \emph{cut score}. 
The idea of the cut score is that a geographically efficient division of units should separate relatively few adjacent pairs across district lines. 
This lets us measure the ``scissors-complexity'' of the division, with the intuition that a plan that deliberately slices up different segments of population in order to meet an agenda will require more scissors-work to cut it into those pieces.

\begin{defn}
A {\em cut edge} of a partition $P$ is an edge of $G$ whose endpoints lie in different districts.  (That is, there is a cut edge for every pair of units that are adjacent to each other but that are assigned to different districts.) 
The {\em cut-set} is the set of all cut edges, and the {\em cut score} counts the cut edges. 
In set-theoretic notation, we can write $$\cut(P):=\left| \Big\{ \{u,v\}\in E : u\in P_i, v\in P_j ~\hbox{\rm for some}~ i\neq j\Big\}\right|.$$ 
\end{defn}

This cut score should be thought of as the discrete analog of the perimeter of the districts:  the boundary length between districts is measured in a count of geographic units, not in miles or kilometers.  This means that perturbations or additional wiggles in the definition of a river, say, do not contribute to the score.  
As an added benefit, only interior perimeter contributes, and not the edge of the state.  
Just as with conventional perimeter, a lower score would be thought of as simpler, so more compact.

To be precise about units, we might speak of the block cut score in a graph dual to census blocks, while Figure~\ref{fig:arkansas} illustrates an Arkansas plan with a county cut score of 36.

The Arkansas county dual graph  has $k=4$ districts, $N=75$ units, and $m=192$ edges.  For all possible graphs with these parameters, the theoretical range of cut edges is from $3$ to   $121$ (see Lemma~\ref{lem:bounds}).  The actual (realizable) minimum number of cut edges possible in a given graph depends on both its connection topology (which vertices are adjacent) and on the population deviation that is allowed, and finding this sharp minimum is closely related to problems that are known to be computationally intractable (i.e., what computer scientists call {\em NP-complete}: if you could solve these problems efficiently, modern cryptography would fall apart).
In this particular case, Becker and Solomon have proved that every possible four-district plan for the Arkansas county graph with $\le 2\%$ population deviation  has at least 32 cut edges, and they produced a plan realizing this bound  \cite[Ch 16]{gerrybook}.   This means that the districting plan shown in the figure, with 36 cuts, is not the most compact possible plan by this measure, but it is fairly close.  

Far from a novel suggestion, using the cut-set as a discrete perimeter is a wholly standard idea across subfields of pure and applied mathematics. Explicit use of cut-sets goes back at least to the Cold War birth of operations research in the mid-1950s  in the form of the Max Flow--Min Cut theorem, establishing a formal and quantitative duality between maximizing  flow in a network and finding a minimal cut-set \cite{FF}.  
A special case was already treated by Menger in the 1920s \cite{menger}.
Graph-based isoperimetric inequalities abound in geometric group theory \cite{BH}, anticipated by important work of Dehn in the 1910s \cite{Dehn} developing graph notions of area and perimeter.
A fairly direct parallel to the usage proposed here is the graph Cheeger constant \cite{lubotzky}, which is computed by finding a short cut that divides a graph into big pieces.  
Connections to network science will be further explored in the next section. As far as we know, the first invocation of the cut-set 
in redistricting was proposed in a political science conference paper by Dube and Clark in 2016 \cite{dube-clark}, where the authors propose to minimize cuts  to optimize ``edge-cut compactness;'' the idea also appears in 
 subsequent work in 2017 \cite{PCD,PDC}. 

The dream of computational redistricting in the 1960s  triggered early use of the graph model for districting plans, because graphs are the mathematical formalism aligned with how computers store adjacency relations.
Today, every computational approach to redistricting makes use of the graph model, and an efficient algorithm for handling partitions is likely to store the cut-set as part of its data pipeline for plans.   This makes the cut score an extremely lightweight calculation in computational applications. The cut score features in  numerous research papers that use Markov chains for redistricting \cite{ReCom,Alaska,competitiveness,VA-criteria,DNS,gerrybook}.  

As a sign of uptake through digital/commercial instantiation, the cut score was incorporated into the updated 2020 release of Maptitude.

\subsection{Spanning trees}\label{sec:spanning-trees}

For the next discrete compactness score, we turn to another aspect of graph theory. This time we build a score of the internal connectivity {\em within} districts, rather than the complexity of the boundary {\em between} districts. 

A graph is \emph{connected} if each vertex can follow a path of edges in the graph to reach any other vertex; otherwise the graph is \emph{disconnected}. 
A connected graph with $n$ vertices must have at least $n-1$ edges. 
When a connected graph with $n$ vertices has exactly $n-1$ edges, it 
is called a {\em tree}; equivalently, a tree is a connected graph with no cycles.  
Every connected graph contains at least one tree as a subgraph that uses all of the vertices, known as a \emph{spanning tree}. In particular, to produce a spanning tree of a graph,  we can proceed by deleting edges, checking not to disconnect the graph at any stage, until no more edges can be removed. Figure~\ref{fig:spanning} gives an example of a graph having four spanning trees.

\begin{figure}[htbp]
\centering
\begin{tikzpicture}[scale=.7]
\begin{scope}[xshift=5cm,yshift=1.75cm]
\draw[thick, rounded corners] (0,0) rectangle (5,3);
\begin{scope}[xshift=.5cm,yshift=.5cm] \draw (0,0)--(3,0)--(3,1); \draw (2,0)--(2,1)--(4,1)--(4,2); 
 \end{scope}
\foreach \x/\y in {0/0,1/0,2/0,3/0,2/1,3/1,4/1,4/2}
{\filldraw (\x+.5,\y+.5) circle (.08);}
\end{scope}

\begin{scope}[xshift=12cm,yshift=3.5cm]
\begin{scope}[xshift=.5cm,yshift=.5cm] \draw (0,0)--(3,0); \draw [dashed] (3,0)--(3,1);
\draw (2,0)--(2,1)--(4,1)--(4,2); 
\draw [line width=3,black,opacity=.3] (0,0)--(3,0);
\draw [line width=3,black,opacity=.3] (2,0)--(2,1)--(4,1)--(4,2);
 \end{scope}
\foreach \x/\y in {0/0,1/0,2/0,3/0,2/1,3/1,4/1,4/2}
{\filldraw (\x+.5,\y+.5) circle (.08);}
\end{scope}

\begin{scope}[xshift=18cm,yshift=3.5cm]
\begin{scope}[xshift=.5cm,yshift=.5cm] \draw (0,0)--(3,0)--(3,1); \draw [dashed] (2,0)--(2,1);
\draw (2,1)--(4,1)--(4,2); 
\draw [line width=3,black,opacity=.3] (0,0)--(3,0)--(3,1);
\draw [line width=3,black,opacity=.3] (2,1)--(4,1)--(4,2);
 \end{scope}
\foreach \x/\y in {0/0,1/0,2/0,3/0,2/1,3/1,4/1,4/2}
{\filldraw (\x+.5,\y+.5) circle (.08);}
\end{scope}

\begin{scope}[xshift=12cm]
\begin{scope}[xshift=.5cm,yshift=.5cm] \draw (0,0)--(2,0);
\draw [dashed] (2,0)--(3,0);
\draw (3,0)--(3,1); \draw (2,0)--(2,1)--(4,1)--(4,2);  
\draw [line width=3,black,opacity=.3] (0,0)--(2,0)--(2,1)--(4,1)--(4,2);
\draw [line width=3,black,opacity=.3] (3,0)--(3,1);
\end{scope}
\foreach \x/\y in {0/0,1/0,2/0,3/0,2/1,3/1,4/1,4/2}
{\filldraw (\x+.5,\y+.5) circle (.08);}
\end{scope}

\begin{scope}[xshift=18cm]
\begin{scope}[xshift=.5cm,yshift=.5cm] \draw (0,0)--(3,0)--(3,1); \draw (2,0)--(2,1);
\draw [dashed] (2,1)--(3,1);
\draw (3,1)--(4,1)--(4,2); 
\draw [line width=3,black,opacity=.3] (0,0)--(3,0)--(3,1)--(4,1)--(4,2);
\draw [line width=3,black,opacity=.3] (2,0)--(2,1);
 \end{scope}
\foreach \x/\y in {0/0,1/0,2/0,3/0,2/1,3/1,4/1,4/2}
{\filldraw (\x+.5,\y+.5) circle (.08);}
\end{scope}
\end{tikzpicture}
\caption{The first image is a graph with $N=8$ vertices and $m=8$ edges.  It has four spanning trees, shown on the right.  They are found by deleting any of the four edges in the graph's lone cycle. }\label{fig:spanning}  
\end{figure}

Since trees are the limiting case of graphs having the most fragile interconnectivity (i.e., no edges to spare), a natural measure of connectedness is the number of spanning trees in a graph.  

\begin{defn} The {\em spanning tree score} of a graph $G$, denoted $\sp(G)$, is defined as the natural logarithm of the number
of spanning trees of $G$.  
The spanning tree score of a districting plan $P=(P_1,\dots,P_k)$ is defined as 
$$\sp(P):=\sum_{i=1}^k \sp(P_i),$$
the sum of the scores of the districts in the plan. \end{defn}

For example, the spanning tree score of the original graph in Figure~\ref{fig:spanning} is $\ln(4)=1.39^-$, because it has four trees. (Here and below, $^+$ and $^-$ will indicate whether the raw value is slightly greater or slightly less, respectively, than the rounded value that is shown.) 
Cut and spanning tree scores for some districting plans in a grid are shown in Figure~\ref{fig:cut-vs-span}.

\begin{figure}[htbp]
\centering
\begin{tikzpicture}[scale=.4]
\def\colone{blue!50}
\def\coltwo{blue!50}
\def\colthree{blue!50}
\def\colfour{blue!50}
\draw[ultra thick] (0,0) rectangle (10,10);
\foreach \x in {1,...,9}
{\draw [gray!50] (\x,0)--(\x,10);}
\foreach \y in {1,...,9}
{\draw [gray!50] (0,\y)--(10,\y);}
\draw[ thick,rounded corners,fill=\colone,fill opacity=.8,draw=black] (0,0) rectangle (5,5); 
\draw[ thick,rounded corners,fill=\coltwo,fill opacity=.8,draw=black] (5,0) rectangle (10,5); 
\draw[ thick,rounded corners,fill=\colthree,fill opacity=.8,draw=black] (0,5) rectangle (5,10); 
\draw[ thick,rounded corners,fill=\colfour,fill opacity=.8,draw=black] (5,5) rectangle (10,10); 
\node at (5,-1.3) [black] {$\cut=20$};
\node at (5,-2.5) [black] {$\sp=80.56^-$};

\begin{scope}[xshift=13cm]
\draw [ultra thick] (0,0) rectangle (10,10);
\foreach \x in {1,...,9}
{\draw [gray!50] (\x,0)--(\x,10);}
\foreach \y in {1,...,9}
{\draw [gray!50] (0,\y)--(10,\y);}
\draw[ thick,rounded corners,fill=\colone,fill opacity=.8,draw=black]
(0,0)--(10,0)--(10,6)--(8,6)--(8,5)--(9,5)--(9,2)--(0,2)--cycle;
\draw[ thick,rounded corners,fill=\colone,fill opacity=.8,draw=black]
(0,8)--(5,8)--(5,7)--(10,7)--(10,10)--(0,10)--cycle;
\draw[ thick,rounded corners,fill=\colone,fill opacity=.8,draw=black]
(0,2)--(4,2)--(4,7)--(5,7)--(5,8)--(0,8)--cycle;
\draw[ thick,rounded corners,fill=\colone,fill opacity=.8,draw=black]
(4,2)--(9,2)--(9,5)--(8,5)--(8,6)--(10,6)--(10,7)--(4,7)--cycle;
\node at (5,-1.3) [black] {$\cut=33$};
\node at (5,-2.5) [black] {$\sp= 65.53^-$};
\end{scope}

\begin{scope}[xshift=26cm]
\draw [ultra thick] (0,0) rectangle (10,10);
\foreach \x in {1,...,9}
{\draw [gray!50] (\x,0)--(\x,10);}
\foreach \y in {1,...,9}
{\draw [gray!50] (0,\y)--(10,\y);}
\draw[ thick,rounded corners,fill=\colone,fill opacity=.8,draw=black]
(7,7)--(3,7)--(3,6)--(5,6)--(5,5)
--(1,5)--(1,3)--(2,3)--(2,1) --(0,1)--(0,0)--(4,0)--(4,1)
--(5,1)--(5,3)--(4,3)--(4,2)--(3,2)--(3,4)--(6,4)--(6,2)--
(7,2)--cycle;
\draw[ thick,rounded corners,fill=\coltwo,fill opacity=.8,draw=black]
(9,9)--(2,9)--(2,8)--(1,8)--(1,7)--(2,7)--(2,6)--(0,6)
--(0,1)--(2,1)--(2,3)--(1,3)--(1,5)--(5,5)--(5,6)--(3,6)--
(3,7)--(5,7)--(5,8)--(8,8)--(8,6)--(9,6)--cycle;
\draw[ thick,rounded corners,fill=\colthree,fill opacity=.8,draw=black]
(0,6)--(0,10)--(10,10)--(10,1)--(9,1)--(9,2)--(8,2)--
(8,3)--(9,3)--(9,4)--(8,4)--(8,5)--(9,5)--(9,9)--(2,9)
--(2,8)--(1,8)--(1,7)--(2,7)--(2,6)--cycle;
\draw[ thick,rounded corners,fill=\colfour,fill opacity=.8,draw=black]
(4,0)--(10,0)--(10,1)--(9,1)--(9,2)--(8,2)--(8,3)--
(9,3)--(9,4)--(8,4)--(8,5)--(9,5)--(9,6)--(8,6)--(8,8)--
(5,8)--(5,7)--(7,7)--(7,2)--(6,2)--(6,4)--(3,4)--(3,2)--
(4,2)--(4,3)--(5,3)--(5,1)--(4,1)--cycle;
\node at (5,-1.3) [black] {$\cut=73$};
\node at (5,-2.5) [black] {$\sp= 14.99^-$}; 
\end{scope}
\end{tikzpicture}
\caption{The $10\times 10$ grid-graph has 100 vertices, 180 edges, and, as a whole, a spanning tree score of $98.45^-$.  
Here we have shown three examples of 4-district plans. 
Both scoring methods rank these model plans from most compact to least compact, reading from left to right---agreeing with the eyeball test.}\label{fig:grid-graphs}
\end{figure}

In the case of spanning trees, a  {\em higher} score represents a more compact plan, because the score measures the internal connectedness of the districts, which visually corresponds to more bulk and fewer bottlenecks. 
The logarithm in the definition helps keep the score in a reasonably human-readable range, since the count itself can get very high once we are in a setting with thousands (or hundreds of thousands) of vertices.
By exploring the spanning tree count of a graph, one soon realizes that ``plumper'' and more interconnected graphs have more spanning trees than their ``spindly'' relatives.\footnote{The precise conditions on graphs (for a given number of vertices and edges) to have as many spanning trees as possible are not known;   this is an open problem even for grid-graphs (those that can be realized in a square lattice) such as those in Figures~\ref{fig:spanning} and \ref{fig:grid-graphs}.
However, it is widely conjectured that square-shaped districts have the most spanning trees in a grid-graph, just as circular districts have the optimal scores for most contour-based measures.}
A thorough discussion of the spanning tree count as a complexity measure, and a beautiful exploration of the asymptotics, can be found in \cite{lyons}.

Spanning tree methods are found across graph theory, theoretical computer science, and electrical engineering.  
Kirchhoff's ``Matrix--Tree'' theorem from 1847 counts spanning trees in a graph as the product of non-zero eigenvalues of the graph Laplacian \cite{kirchhoff}.
Wilson's algorithm samples a spanning tree uniformly at random, using a random walk method \cite{wilson}.  Kruskal's algorithm (a favorite in undergraduate computer science courses) gives a faster method to generate spanning trees in a weighted graph using random edge weights \cite{kruskal}.  
Network theory and practical network protocols make heavy use of spanning trees.
The use of spanning trees for clustering is discussed further in the next section.


\section{Interpreting and assessing discrete compactness}\label{sec:assess}

This paper calls for importing  ideas from discrete geometry into the study of electoral geography, so we need broader context to situate these mathematical ideas, setting up an evaluation of the fit.  
Agnew and Duncan called for geographers to ``display more critical acuity in borrowing ideas from outside the field'' \cite{agnew-duncan}; 
the mere fact that cuts and spanning trees appear all over the mathematics and computer science literature is no guarantee that they help us think about compactness specifically, or human geography more broadly.
In this section we take on the conceptual work of connecting metric to meaning, building an argument that these scores are well suited to measure compactness in political geography terms, and that they may be promising for wider application.  For the specific use case of redistricting, we will consider how the new scores hold up against the problems and issues outlined above.

\subsection{Clusters and communities}\label{sec:interp}

Both the cut score and the spanning tree score reward plans that keep well-connected areas together, and both prefer plans that find economical  rather than winding cuts.  This means that both scores will tend to penalize districts that weave or meander through space, but importantly this is measured in terms of the passage from unit to unit and not as the crow flies.  They maintain attention on the building blocks of districting plans and study them as a network in which {\em adjacency} matters, not Euclidean distance or travel patterns or demographic similarity (except in the ways that those attributes are reflected in the units themselves and their connection topology).  This will capture interesting structure when the geographical units are built to correspond to human political and social formations.\footnote{Notably, a government office exists to build and uphold concepts and criteria for many of the units discussed here.
The Geography Division of the U.S. Census Bureau   maintains a Geographic Standards
and Criteria Branch  whose publications frequently reference social structure in the definition of geographies like census blocks and block groups (which are built and adjusted in collaboration with local officials) \cite{concepts}.  This institutionalized attention to concept maintenance is notable, though of course it provides no guarantee of correspondence to the lived experience of residents.}

The notion of selecting subgraphs with high interconnectivity is well studied in the field of network science, going by the highly suggestive name of 
{\em community detection}.  As one survey article puts it,
``One mesoscopic structure, called a {\em community}, consists of a group of vertices that are relatively
densely connected to each other but sparsely connected to other dense groups in the network'' \cite{POM}. 
The use of community structure language and methods from network science is well underway in the geography literature (see \cite{adam-etal,chi-etal,he-etal,thomas-etal}).
This dovetails with the different but highly related study of {\em clustering} methods in statistics (and now in machine learning), for which spanning tree algorithms were being used at least as far back as 1969 \cite{GR,zahn}.  For the uses of clustering in  regionalization, see for instance \cite{farmer-foth}, which uses terms that closely echo the networks notion of community from above:  a functional region is ``internally well connected and relatively cohesive, especially compared with the links between regions.''

So favorable cut and spanning tree scores occur when a plan offers an efficient graph clustering with weight balance.
It is important not to overstate the congruence of the graph-theoretic definition of community with the more robust social and cultural definition; the correspondence is more than cosmetic but far less than complete.

\subsection{Comparison, normalization, and bounds}\label{sec:compare}
\begin{figure}[htbp]
\centering
\begin{tikzpicture}
\node at (0,0) {\includegraphics[height=1.6in]{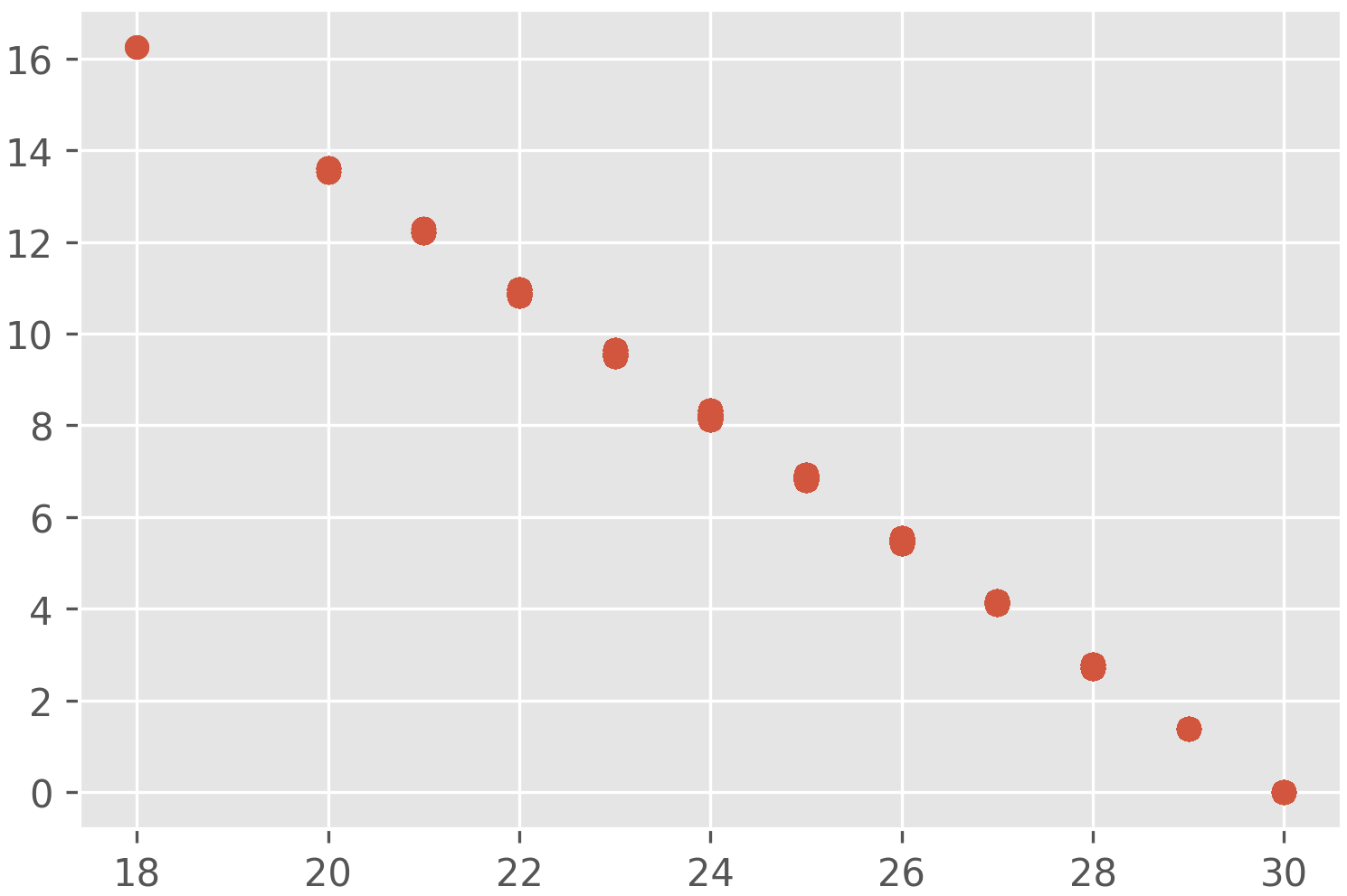}};
\node at (0,-2.5) {$\cut(P)$};
\node at (-3.5,0) [rotate=90] {$\sp(P)$};
\node at (8,0) {\includegraphics[height=1.6in]{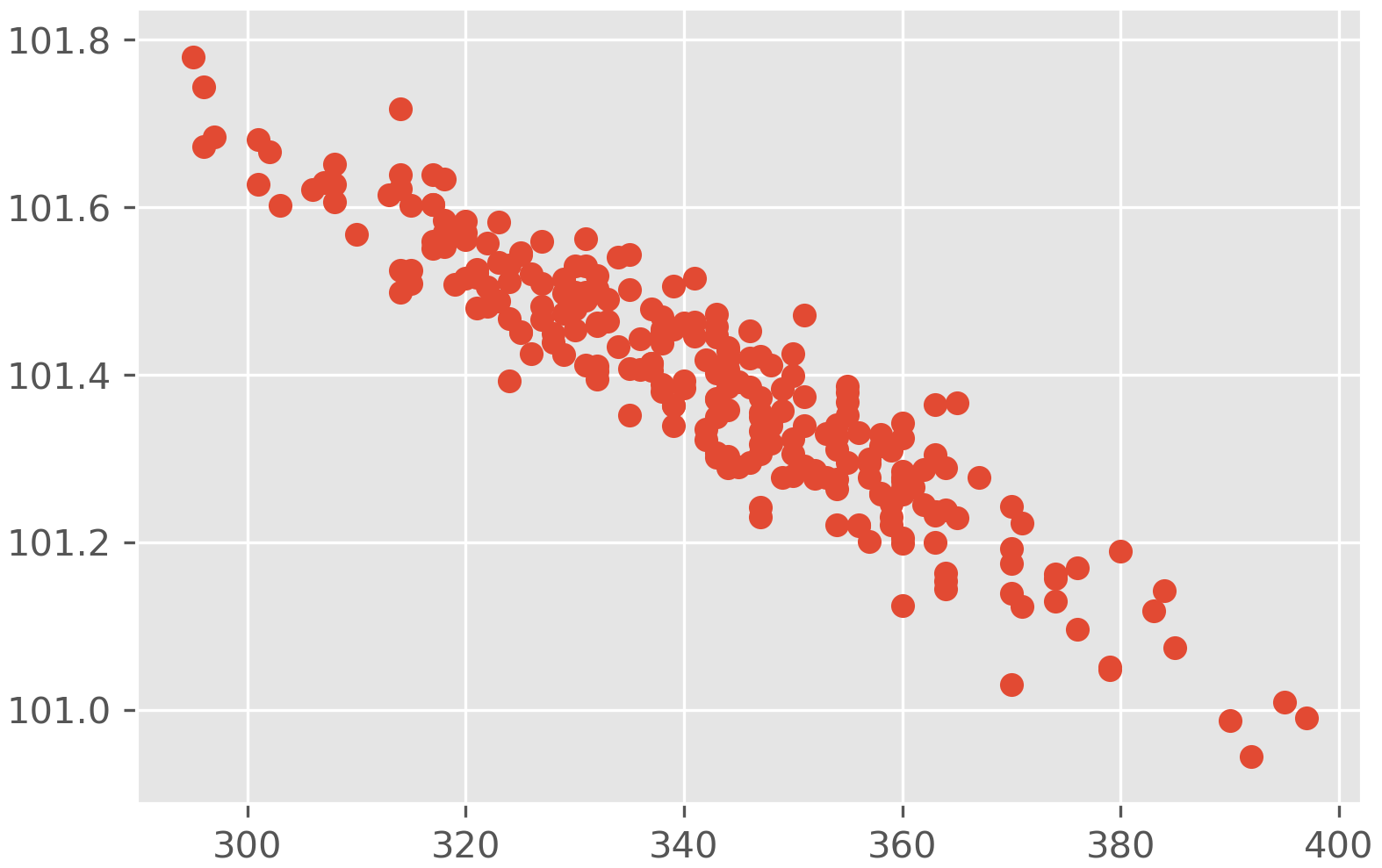}};
\node at (8,-2.5) {$\cut(P)$};
\node at (4.35,0) [rotate=90] {$\sp(P)$};
\end{tikzpicture}
\caption{These charts present comparisons of $\cut$ and $\sp$ scores. On the left is a scatterplot of all 451,206 six-district plans on the $6\times 6$ grid, showing a nearly perfect linear relationship between the scores.  On the right is the  same comparison for a  random sample of 250 40-district plans on a $40\times 40$ grid, showing a strong negative correlation. These were sub-sampled from an ensemble of 100 million plans generated by a Markov chain process described in \cite{ReCom}, using an implementation available in \cite{GerryChain}.
The latter problem, building 40 districts from 1600 geographical units, has enough complexity to be comparable to a real-world redistricting problem.}\label{fig:cut-vs-span}
\end{figure}

The cut score and the spanning tree score are strongly related to each other (see Figure~\ref{fig:cut-vs-span}), and are generally correlated with the standard suite of scores (Polsby-Popper, Reock, etc.) when applied to enacted districts.\footnote{It is easy, however, to design illustrative examples where the contour scores diverge from the discrete scores---and from each other---in their assessment of whether districts are compact or not.}  But we will argue in the next section that these discrete scores mitigate (though they do not eliminate) the worst issues and limitations of the classical scores.  

First, we briefly address the question of normalization:  is it possible, or advisable, to transform these scores to be within a fixed range (like 0 to 1) for the purposes of comparison?  To address this we should say something about the highest and lowest possible scores.  The lowest possible spanning tree score is $0$, which occurs when every district is a tree.  We can summarize what is known about the other bounds as follows.

\begin{lemma}[Bounding the cut and spanning tree scores]\label{lem:bounds}
Suppose $G$ is any connected planar graph with $N$ vertices and  $m$ edges.
Over all possible partitions $P$ into $k\ge 2$ connected districts, the cut and spanning tree scores satisfy the following inequalities:
$$k-1 \le \cut(P) \le m-N+k \ ; \qquad  0 \le \sp(P) \prec 1.665N.$$
Spanning tree scores of roughly $1.615N$ are realized by families of partitions on a triangular lattice.
\end{lemma}

The notation $\prec$ represents an asymptotic inequality:  that is,  $\sp(P)\prec 1.665N$ means that $\sp(P)$ is much smaller than $1.665N$ when $N$ is large.  For more details and the proof of Lemma~\ref{lem:bounds}, see Appendix~\ref{sec:lemma}.

Considering these bounds, is it possible to use discrete compactness scores to compare two plans made from different graphs, such as a congressional plan in Wisconsin against one from North Carolina?  
We emphasize that the bounds in Lemma~\ref{lem:bounds} hold for all graphs with $N$ vertices and $m$ edges, including extreme examples (like paths) that are very unlike geography dual graphs.  Finding a precise maximum and minimum achievable score for a graph in a given class is a difficult open problem.  But of course, the same is true for the classical scores; even though $1$ is the best Polsby-Popper score for a single district in the abstract, that value is only  achievable if the boundary is a perfect circle.  Circular districts are geographically unrealistic, in the first place, and would not fit together to tile a region in any case.  The best Polsby-Popper score for a real-world redistricting problem of finding equipopulous partitions of census geography seems to be just as difficult to assess.

So instead of attempting to standardize the range of scores to encourage cross-context comparison, we argue that the (un-normalized) cut and spanning tree scores promote the good practice of only comparing plans to their alternatives that hold constant the problem's parameters (the choice of region, building blocks, and number of districts).  
That is, the practice of normalization and cross-state comparison was always misleading, even for standard compactness scores.

\subsection{Assessing success}\label{sec:discrete-assess}

We have argued that the redistricting data is better suited to discrete geometry; now we will consider whether the issues discussed in \S\ref{sec:issues}, which attach to contour-based scores of compactness, are successfully defused by discrete scores like $\cut$ and $\sp$.  

\begin{figure}[htbp]
\centering
\includegraphics[height=2.4in]{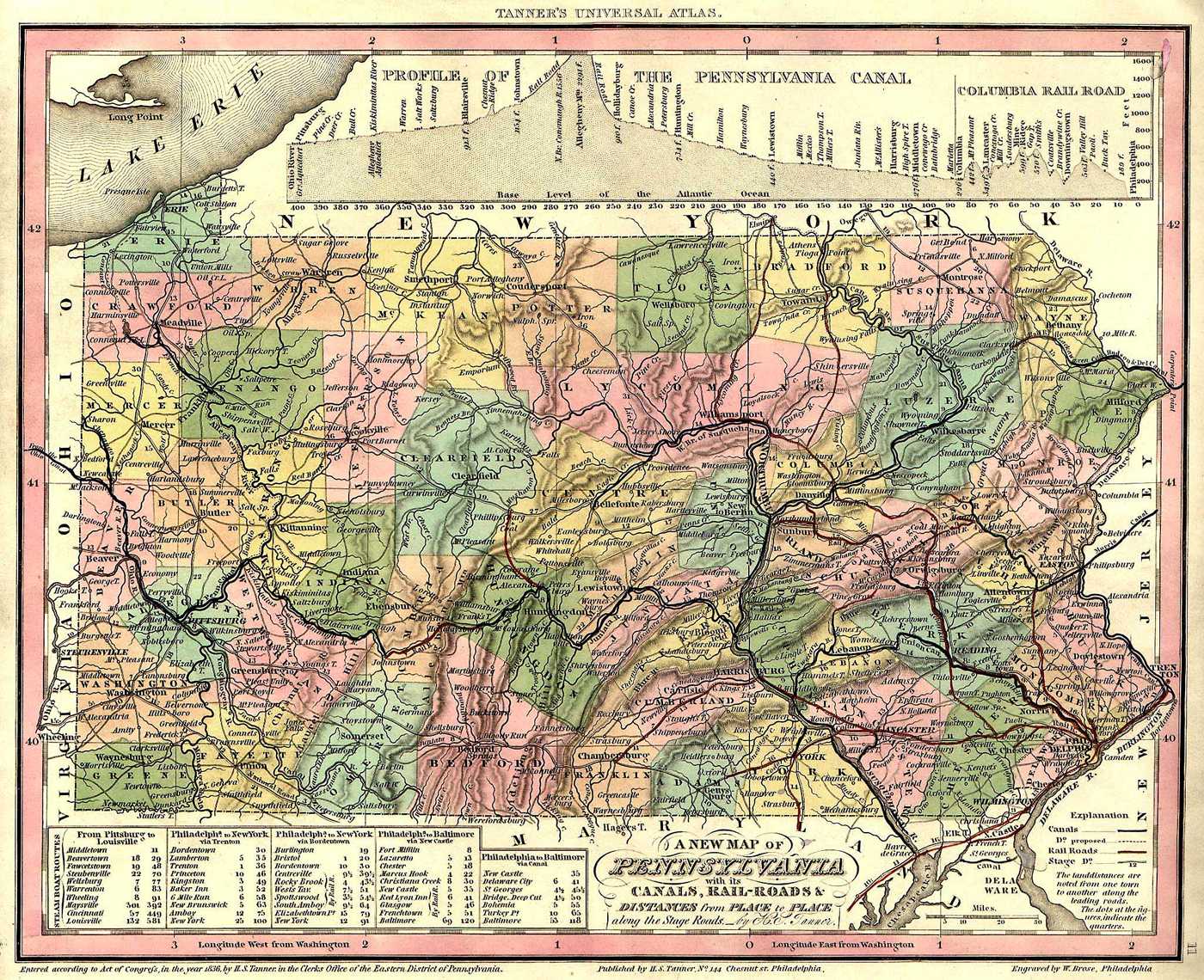}  \quad \includegraphics[height=2.4in]{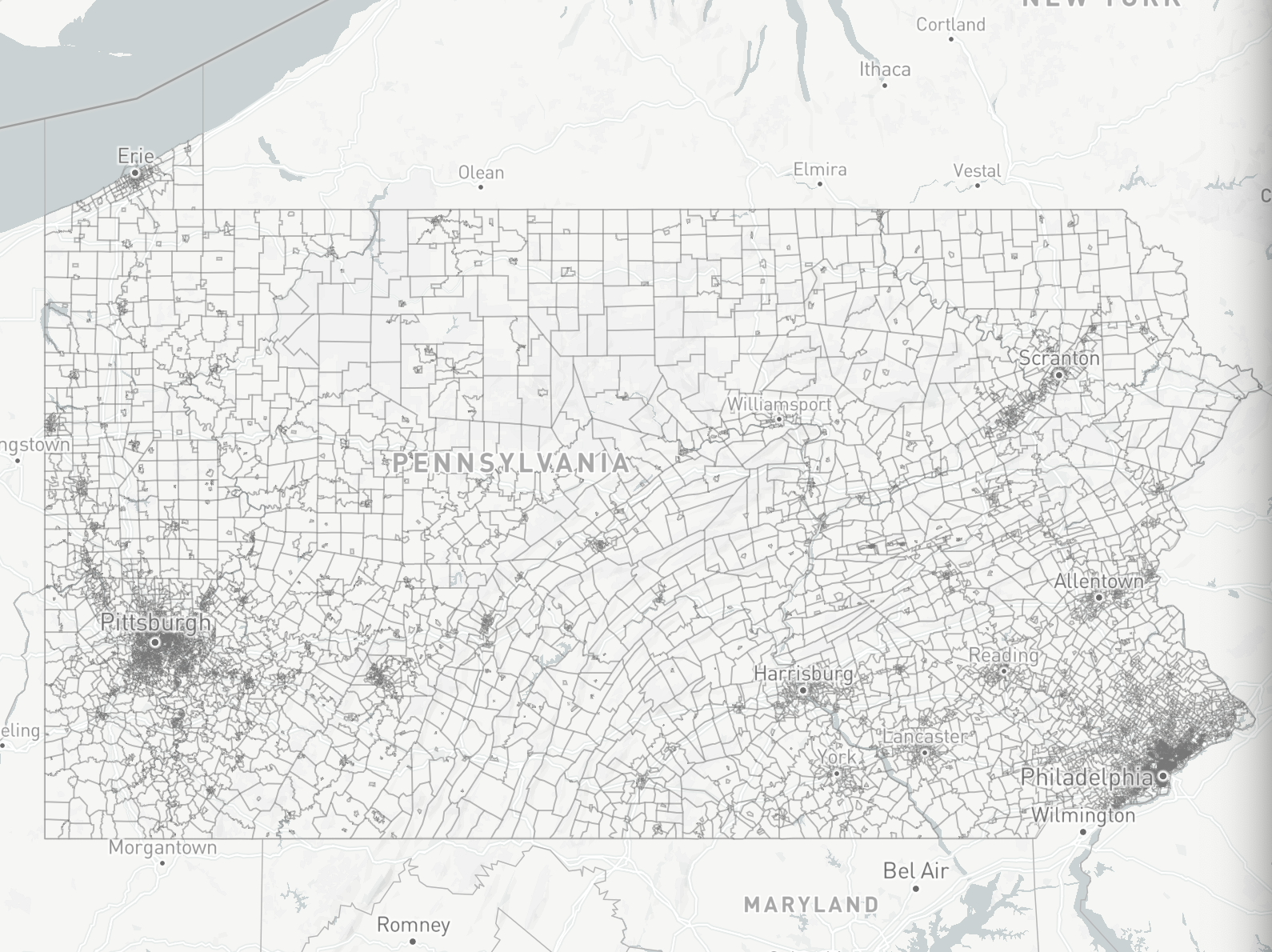} 
\caption{A comparison of physical geography (left) to precinct geometry (right) in Pennsylvania. We see that precincts are not a lattice superimposed on the state, but instead reflect patterns of natural geographical formations and built infrastructure.}\label{fig:PA-geog}
\end{figure}

Discretization will significantly mitigate coastline effects (Issue A), 
since the census geography along a complicated natural border will absorb any wiggling perimeter into a fixed (and not unduly large) number of units. Subsequent refinements to the measurements that add precision to those polygons will not increase the number of polygons, and so will not change the discrete scores---the geographical units encountered along a jagged coastline will be more complicated, but not generally more {\em numerous}, than along a straight-line boundary.  
Since the discrete scores are defined by counts over the units and their adjacencies, they do not incur a coastline tax.
More broadly, it is an interesting question to consider the extent to which physical geography is reflected in census and administrative units (Figure~\ref{fig:PA-geog}).

Some insulation from resolution instability (Issue B) is provided by 
discrete compactness scores for the same reason:  more precision puts more points on each polygon, but typically without changing the adjacencies between units.  

\begin{figure}[htbp]
\centerline{\includegraphics[width=4.5in]{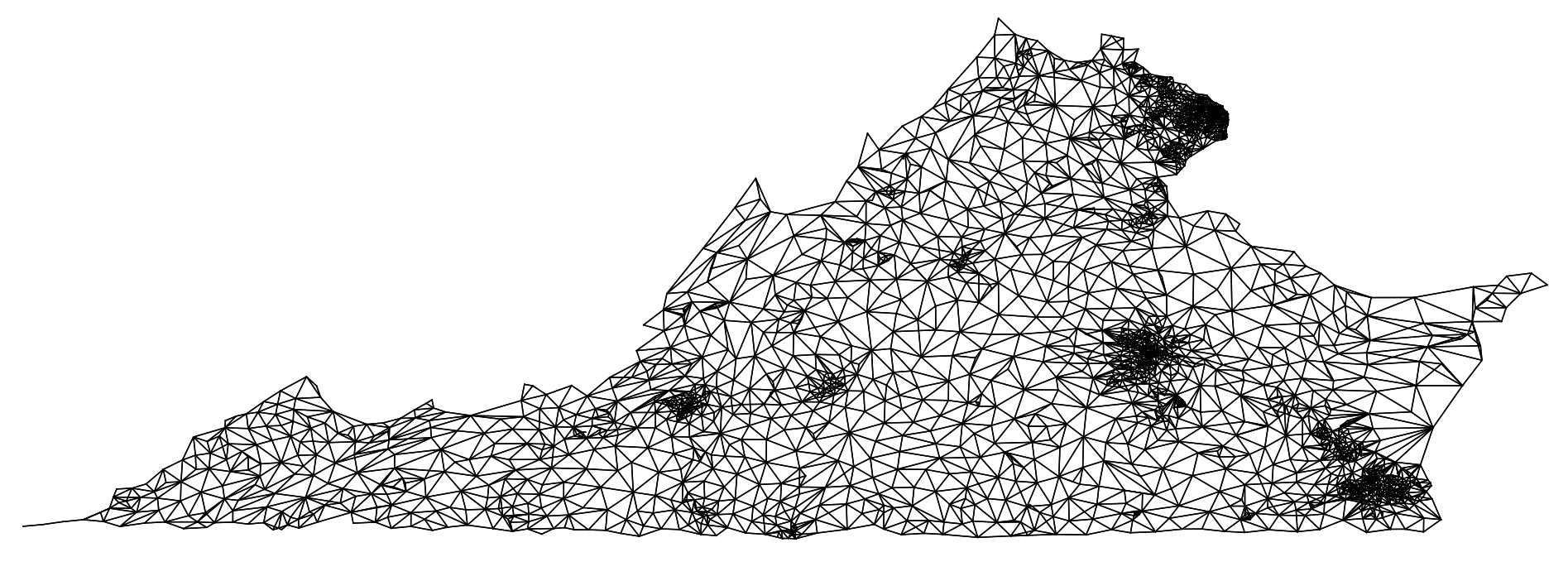}}
\caption{The dual graph for the  $2372$ VTDs (precincts) in Virginia, shown with each vertex plotted at the centroid of the VTD it represents.  The corresponding
graph of the $285,\!762$ census blocks  would look substantially more complicated.
The centroidal embedding makes cities visible as loci of greater vertex density, but the graph itself is just a list of vertices and a record of which pairs of vertices are adjacent. \label{fig:virginia}}
\end{figure}

Next, note that coordinate data is not recorded in the graph from which our discrete scores are computed.
Figure~\ref{fig:virginia} shows a centroidal embedding of the VTD graph in the plane for illustrative purposes, but the weighted graph itself retains no latitude/longitude data, only adjacency relations and population counts for the units.
This means that any compactness score based on the graphs dual to geographical units---in contrast with all contour-based scores---will automatically be independent of map projection or choice of coordinates. This completely negates Issue C above.

Finally, discrete compactness scores that do not  take population weights into account, like the cut and spanning tree scores in the form presented above, are still subject to empty space effects (Issue D).   
In particular, since so many census blocks have zero population, their movement between districts creates a different partition (and changes the $\cut$ and $\sp$ scores) but not in a way that affects  representation.\footnote{In order to ensure that scores are more keyed to population than area, one approach is to work with a graph based on units that are designed to have relatively even levels of population, such as block groups or precincts. The Census Bureau writes that block groups are ``statistical subdivisions of census tracts[,] generally defined to contain between 600 and 3,000 people and 240 and 1,200 housing units'' 
\cite{glossary}.} 
The desire to put more emphasis on population might lead us to favor population-weighted scores, 
which are designed not to ``see'' unpopulated areas.  
For instance, instead of reporting the number of cut edges, we can sum over the populations of their endpoints, which 
incentivizes districts whose borders slice  through  sparsely populated areas. This may be in line with good districting practices in some instances---such as keeping together a community of interest---but at odds in other instances, because city populations are vulnerable to packing and are sometimes better served by being split.   

Taken together, this means that the area-vs.-population tension is the issue that remains most alive for discrete compactness scores, while the other problems are significantly mitigated.  
However, this tension may not be resolvable.  It is area that governs the eyeball test, and so area has an ineliminable role in the judgments that will ultimately be passed by the public, the bodies that control the lines, and the judges who decide their fate when challenged \cite{king-ML, bizarre}.  

Let us  review  the design principles for districts that have been written into the rules and applied by courts and by observers.  Compactness has been hoped to flag divisions with excessive geographical complexity, to limit the power of line-drawers, and to promote cognizability and ease of description. 
The cut and spanning tree scores favor districts that pass muster visually but also, by construction, are formed as clusters in the adjacency network of census or administrative units.  
Placing limitations on the discrete compactness scores allowed in districts will restrict the degrees of freedom of the redistricter just as contour scores do.\footnote{Civil rights advocates have been leery of compactness scores, worrying that they could obstruct the formation of effective districts under the Voting Rights Act.  But this worry can be turned around: because the cut score in particular is so efficient to compute, algorithmic methods can be used to find plans with VRA-compliant districts and reasonable compactness, as in \cite{BDGH}.  Indeed, the opinion of the court in {\em Milligan}, authored by Chief Justice John Roberts, favorably cites just this use of exploratory algorithms. (See footnote 7 of that decision.)}
And by mitigating the coastline tax and  measuring complexity in units that are themselves built to hew to boundaries of towns and relevant physical geography, discrete compactness scores are better aligned with the formation of cognizable districts.  Indeed, a plan can be completely described by specifying its units and naming its cut edges.  So this is a final and very literal sense in which plans with fewer cut edges are more simply described.

\subsection{Empirical benchmarks and directions for future work}\label{sec:benchmarks}

We close by sketching future directions for quantifying compactness.

\subsubsection*{Other scores}

We have focused attention here on two possible discrete compactness scores, but there are an enormous number of possibilities for the application of discrete ideas to redistricting.  The cut score is based on a discretization of perimeter.  A corresponding discretization of area would open the door to discrete isoperimetry. (A discrete Polsby-Popper score is defined in Appendix~\ref{sec:isoperimetry}.)  

We have focused on the limitations of plane geometry, but there is another interesting possible departure from that way of measuring:  distance calculated in miles can be replaced with distance calculated in travel time, as numerous authors have noted.  
For example, dispersion-based scores have been formulated for decades that are essentially average distance calculations over districts, say the average distance between pairs of points or from a point to the center of mass.  (See the ``moment of inertia'' scores in \cite{NGCH}.) 
These can first be discretized, localizing population by census block and using sums over the blocks rather than integrals over the plane domains.  
If distances are then computed with respect to travel time rather than linear miles, we might find that time compactness---low ``travel-time dispersion"---tracks with easy transit and assembly.  This sounds quite appealing.  What makes this challenging to execute is the lack of a canonical or stable source for travel times that parallels the census as a canonical source for geographical units.  Even more problematically, to the extent that travel time is supposed to be human-centered, we have to ask whose travel we are timing:  residents with cars or who can pay for fast trains will have very different travel realities from those who rely on buses, even for the same routes.

\subsubsection*{Units}
The choice of units for analyzing plans is a thorny question, tied to the equally thorny choice of units for building plans in the first place.
We have argued census blocks are in many ways a good choice because current practice is to tune plans at that level, but building with larger units would in many cases be a good practice (few choices, more cognizable, and so on).
When trying to define compactness scores using larger units that do not cleanly nest in districts, one strategy is to refine the units by splitting down to their intersections with the districts, but this has the downside that it becomes hard to compare a given plan with other alternatives on equal footing.  For the cut score in particular, a quantitatively equivalent approach is to increment the cut edge count for each split unit, effectively regarding it as made up of multiple sub-units that have been severed from each other in the plan.  
A second strategy is to ``round off" the districting plan into a given set of units, by assigning each unit to the district with the largest share of its population, say.
Our preliminary explorations indicate that all of these strategies retain the motivating properties of discrete scores when working with precincts for congressional and state legislative districts, and sometimes block groups, but that census tracts are too large (and too independent of district structure)---attempts to describe districts in terms of tracts will lose the signal of unit integrity that discrete compactness scores are designed to detect.  There are exceptions to this rule of thumb that precincts are the largest reasonable choice:  for instance, the practice in Iowa is that congressional districts are built from whole counties.  An individualized inquiry should be conducted when choosing suitable units for undertaking a new kind of redistricting analysis with discrete geometry tools.

\subsubsection*{Incentives and gaming}
Sometimes, scores that are built with reference to one set of inputs turn out to be sensitive to other features, in addition to (or sometimes even instead of) the attributes appealed to in the definition.
This occurs so commonly in redistricting metrics that it is the rule rather than the exception:  from {\em efficiency gap} to {\em competitiveness} to {\em partisan symmetry}, many popular scores turn out to have surprising simplifications that show they can't possibly capture all the phenomena that their proponents used to motivate and justify them \cite{cover,veomett, competitiveness,PSymm}.
In some cases these findings can be worked out directly by analytical examination of the definitions, and in some cases there are suggestive correlations that can be observed from computational investigation.  In fact, observed correlations can sometimes lead to the derivation of provable characterizations.\footnote{This was the case with the ``Utah paradox'' in \cite{PSymm}.}

It is important to study the correlates of good or bad performance in any new proposed score, in order to be confident that irrelevant plan features are not being unduly rewarded or penalized, and that relevant features move the needle.
More generally we should seek to be sure that the named inputs do not carry hidden proxies.  
In particular, all metrics that are designed to evaluate districts (and all rules and generation methods for plans)  merit a close look at the incentives they create for the line-drawers.  An example already came up just above:  a population-weighted perimeter might be intended to mitigate coastline effects, but by neglecting unpopulated units it will rate plans as efficient, or compact, when they slice through rural or industrial areas and leave more densely populated regions whole.  This might be advantageous in some circumstances but could not be expected to produce a healthy incentive overall.

The flip side of incentives is the gaming of scores, where agenda-driven plans are designed to avoid detection by finding blind spots or loopholes in the gatekeeping metrics.  Following through with the example of a population-weighted perimeter,   
an abusively drawn district could, in principle, make its gerrymander invisible by adding a buffer of unpopulated units around its border, thereby dropping the weighted perimeter to near zero.

To address both of these, an important method of approach is to build large ensembles of plans to examine, looking for correlations between discrete compactness scores and other map features, both under neutral conditions and while searching for extreme examples.  
A systematic correlational study of new compactness metrics  will also help clarify the success with which mathematized notions of communities and geographical clusters map onto bottom-up social understandings of communities of interest. 

\bigskip

Overall, this discussion leaves us with two promising measurements of compactness that are cued up for investigation. 
We have argued that these metrics go further to meet the normative aspirations of compactness scores for redistricting than the classical scores.
At the same time, there is ample room for further computational (and normative) exploration  of applications of discrete geometry to redistricting.
We hope that political scientists, sociologists, legal scholars,  mathematicians, and political geographers will find this to provide intriguing questions for study, working in parallel and (especially) in collaboration.

\newpage

\appendix

\section{Ideas for discrete isoperimetry}\label{sec:isoperimetry}

Many geometric ideas have been profitably discretized in the last several decades, particularly in computer science, where discrete differential geometry is an essential part of computer graphics, and geometric group theory, where the shapes of spaces are echoed in the shapes of their (discrete) lattices (see generally \cite{DDG,BH,OHGGT}.  Consequently, another natural direction of inquiry is to take the popular definitions of compactness and to discretize their elements, like area and perimeter.  In this appendix we will briefly outline the ideas needed to form a discrete Polsby-Popper score.

Recall that the weight  of the vertices in the dual graph $G$ is given by the  population of the corresponding units.  
We can define the boundary $\partial G$ of the graph to be the subset of vertices corresponding to units on the outer boundary of the state---note that this is not an abstract graph notion, but depends on the tiling from which the dual graph was made.  Consider a partition $P=(P_1,\dots,P_k)$ into $k$ connected districts.  Then we can similarly define the boundary
$\partial P_i$ of each district coming from units on the outer boundary of the district (either belonging to $\partial G$ or adjacent to a different district).

Let us define the 
{\em discrete area} of a district to be the order (that is, the number of vertices) of the  district and the 
{\em discrete perimeter} to be the order of its boundary, possibly choosing to weight 
both of these calculations by population.
This immediately suggests two discrete analogs of the Polsby-Popper score of a district.
We can let the compactness be measured by
\begin{enumerate}
\item\label{raw-ratio} discrete area divided by the square of discrete perimeter; or 
\item\label{weighted-ratio} the same calculation, but weighted by population.
\end{enumerate}
These scores are, respectively,
$$
\hbox{\rm (1)} \quad \frac{|\Omega|}{|\partial\Omega|^2} \qquad \hbox{\rm and} \qquad \hbox{\rm (2)} \quad
\frac{\sum\limits_{v\in\Omega} w(v)}{\left(\sum\limits_{v\in\partial\Omega} w(v)\right)^2}.$$

To defend the decision to square the perimeter to achieve scale-independence, as in the continuous setting, consider the lattice examples in Figure~\ref{fig:lattices}.

\begin{figure}[htbp]
\begin{center}\begin{tikzpicture}[scale=.4]
\foreach \layer in {1,...,5}
\foreach \x in {-\layer,...,\layer}
  {  \draw (\x,-\layer-.6)--(\x,\layer+.6);  \draw (-\layer-.6,\x)--(\layer+.6,\x);
  \foreach \y in {-\layer,...,\layer}
  	{\filldraw (\x,\y) circle (.1);
	}}
\foreach \x in {-3,...,3}
  {
  \foreach \y in {-3,...,3}
    {\draw[fill=red, fill opacity=.5] (\x,\y) circle (.3);}}	

\begin{scope}[xshift=18cm]
\def\sqt{.866}
\filldraw (0,0) circle (.1);
\draw  [fill=red, fill opacity=.5]  (0,0) circle (.3);

\def\level{6}
\clip (0:\level+.6)--(60:\level+.6)--(120:\level+.6)--(180:\level+.6)--(240:\level+.6)--(300:\level+.6)--cycle;
\foreach\tick in {0,60,...,300}
{\begin{scope}[rotate=\tick]
\draw (-\level,0)--(\level,0);
\foreach \layer in {1,...,\level}
   { \draw (-\level,\layer*\sqt)--(\level,\layer*\sqt);
   \foreach \k in {1,..., \layer}
       {\filldraw (\k-\layer/2,\layer*\sqt) circle (.1);}}
\end{scope}}

\def\level{3}
\foreach\tick in {0,60,...,300}
{\begin{scope}[rotate=\tick]
\foreach \layer in {1,...,\level}
   {    \foreach \k in {1,..., \layer}
       {\draw [fill=red, fill opacity=.5] (\k-\layer/2,\layer*\sqt) circle (.3);}}
\end{scope}}
\end{scope}    
\end{tikzpicture}
\end{center}
\caption{Square and triangular lattices with square and hexagonal "districts" $\Omega_7$ and $\Omega'_4$, respectively.\label{fig:square lattice}\label{fig:lattices}}
\end{figure}

The square-shaped subgraph $\Omega_n$ in the square lattice $G$ 
and the hexagon-shaped subgraph $\Omega_n'$ in the triangular lattice $G'$ have 
isoperimetric ratios 
$$\frac{|\Omega_n|}{|\partial\Omega_n|^2}= \frac{n^2}{16(n-1)^2}  \hspace{.25in} \text{and} \hspace{.25in}
\frac{|\Omega_n'|}{|\partial\Omega_n'|^2}= \frac{3n^2-3n+1}{36(n-1)^2},$$ respectively. 
These 
tend to positive, finite limits as $n$ gets large ($1/16$ for the square case and $1/12$ for the hexagon), whereas if any other 
power of perimeter had been used, the limits would be zero or infinity.  
We interpret this to say that if a grid has underlying geometry that is roughly Euclidean, then 
squaring the perimeter makes these measurements stable under refinement of the grid.

There is no need for a coefficient in the discrete calculation to play the role of $4\pi$ from the classical formula.
This is because $4\pi$ was chosen in order to scale the continuous value $\PP$ to lie in the unit interval, whereas these discrete variants can take arbitrarily large or small values.\footnote{With a uniform weighting function, a path of length $n$ would have isoperimetric ratio approaching zero; on the other hand, we can produce scores
tending to infinity if we begin with a fixed district $\Omega$ and successively subdivide interior cells while leaving the boundary fixed.} We consider this to be a feature, not a bug:  it reminds the responsible modeler to only 
compare a compactness score to others that have been collected at the same resolution---which is a good practice,
whether scores are contour-based or discrete.

As a remark, the discrete geometry explored here---small cut-sets, 
interconnected clusters, and isoperimetric inequalities---can be rephrased and reframed in terms of discrete curvature.  Graphs that encode geographical networks carry their own geometry.  In picturesque words, imagining the edges as rigid rods with length one, for instance,  allows us to convert the connection topology to a surface topography with peaks and saddles and plateaus.  At a high level, discrete definitions of compactness measure how the divisions cleave this abstract landscape as much as the physical one.  

Finally, we make a quick note on allocation of split units.
When the units represented by a dual graph nest neatly into districts, then it is straightforward to regard the plan as a partition:  we let $P_1$ be the subgraph corresponding to the units in district 1, $P_2$ correspond to district 2, and so on.  
However, we sometimes want to build a graph from units that do not fully nest in districts.
As we have seen, census block inclusion in districts is usually all or nothing, but
 districts typically split the larger census units.  
Therefore, if bigger units are being used,   an allocation system is needed in order to decide which 
units belong to which districts.
For example, the modeler might assign each unit
to the district in which the largest part of its land area lies, or the largest portion of its population. 
Alternatively, the units themselves can be split to make smaller pieces that do nest, with fractional allocation of the original population to each district, assuming the share of population in each district can be estimated.

We have formulated definitions of compactness for districting plans that make use of the formalism of graph partitions; the impact of making different allocation choices should be studied for any proposed metric if blocks are not the basic unit.

\section{Bounding the discrete scores}\label{sec:lemma}

We now prove Lemma~\ref{lem:bounds}.  Below we 
write $f(n)\asymp g(n)$ and $f(n) \prec h(n)$ if the limiting behavior of a function $f$
satisfies $\lim_{n\to\infty} f(n)/g(n) =1$ and $\lim_{n\to\infty} f(n)/h(n) =0$, respectively.

\setcounter{thm}{4}
\begin{lemma}[Bounding the cut and spanning tree scores]
Suppose $G$ is a connected planar graph with $N$ vertices and  $m$ edges.
Over all possible graphs $G$ and all possible partitions $P$ into $k\ge 2$ connected districts, the cut and spanning tree scores satisfy the following inequalities:
$$k-1 \le \cut(P) \le m-N+k \ ; \qquad  0 \le \sp(P) \prec 1.665N.$$
Spanning tree scores of roughly $1.615N$ are realized by families of partitions on the triangular lattice.
\end{lemma}

Before we begin, we cite spanning tree counts for planar lattices and a global bound for planar graphs.  Keeping the notation from   Figure~\ref{fig:lattices}, we can write $\Omega_n$ for a square subgraph of the square lattice, and 
$\Omega_n'$ for a hexagonal subgraph of the triangular lattice.  
Classical counting formulas  tell us that  
$$\sp(\Omega_n)\asymp 1.166^+ N \qquad \hbox{\rm and} \qquad \sp(\Omega_n')\asymp 1.615^+ N,$$
where $N$ is the number of vertices of each graph: $|\Omega_n|=n^2$ and $|\Omega_n'|=3n^2-3n+1$ \cite{GW}.
In a 2010 paper, Buchin and Schultz use a linear programming argument to establish that all families of planar graphs satisfy 
$\sp(G_n)\prec 1.665^-N$ for $N=|G_n|$ \cite{BS-num}.

\begin{proof}
Because trees are minimally connected, every deleted edge divides them into two connected components; for any other graph, each cut edge may or may not separate the graph.  Therefore it takes $k-1$ cut edges to divide a tree into $k$ connected components, and it takes at least this many for general graphs.

At the other extreme, we construct a graph with a maximal cut score relative to its total number of edges.  This occurs when the districts themselves have the lowest number of edges.  Suppose the districts $P_1,\dots,P_k$ have $n_1,\dots,n_k$ vertices, so that $n_1+\dots+n_k=N$.  Since each district must be connected, having $n_i$ vertices requires at least $n_i-1$ edges.  This means that there are at least $\sum_{i=1}^k (n_i-1)=\left(\sum n_i\right)-k=N-k$ edges within the districts, leaving $m-(N-k)=m-N+k$ edges of $G$ that were cut to obtain this partition.

Consider again a partition into trees.  A tree $T$ has only itself as a spanning tree, so  $\sp(T)=\ln(1)=0$.  That means that a plan whose districts are trees satisfies $\sp(P)=0+\dots+0=0$.

Finally, let us denote by $N_{\sf ST}(G)$ the number of spanning trees of a graph $G$.  Using the definition of $\sp$ and the properties of logarithms, we have 
$$\sp(P)=\sum_i \ln(N_{\sf ST}(P_i) )= \ln\bigl( N_{\sf ST}(P_1) \cdots N_{\sf ST}(P_k)\bigr).$$
The product of the number of spanning trees of the districts counts the number of ways to simultaneously choose one spanning tree for each.
But any choice of spanning trees for the districts can be completed to a spanning tree of the full graph $G$ by restoring cut edges between districts (and there must be enough cut edges between districts for $G$ to have been connected in the first place).  This gives us $\prod_i N_{\sf ST}(P_i)\le N_{\sf ST}(G)$, which shows that  $\sp(P)\le \ln(N_{\sf ST}(G))=\sp(G)$.

Finally we establish the claim about a family of partitions of triangular lattice graphs.  Consider a graph 
$G$ partitioned into $k$ copies of $\Omega_n'$ for large $n$, writing $N$ for the number of vertices of $G$ so that 
$|\Omega_n'|=N/k$.   Then 
$\sp(P)=k\cdot \sp(\Omega_n') \asymp  k \cdot c (N/k) =cN$, as desired, where $c=1.615^+$.
\end{proof}

As to finding planar graphs with many spanning trees, we note that any planar embedding that is a contender for maximality would have to be triangulated, because otherwise we can increase the number of spanning trees by adding a diagonal to some face.  The standard triangular lattice has degree six at each vertex; we can do slightly better with some vertices of higher degree, but get diminishing returns as we push up the degree.  

\end{document}